\documentclass{aa}

\usepackage[colorlinks=true, linkcolor=blue, citecolor=blue, urlcolor=blue]{hyperref}
\usepackage{txfonts}
\usepackage{threeparttable}
\usepackage[table,xcdraw]{xcolor}
\usepackage{longtable}
\usepackage{booktabs}
\usepackage{multirow}
\usepackage{geometry}
\usepackage{threeparttable}
\usepackage{graphicx}
\usepackage[utf8]{inputenc}
\usepackage{newunicodechar}
\newunicodechar{χ}{\chi}
\usepackage{gensymb}
\usepackage{tabularx,booktabs,makecell}
\usepackage{enumitem}
\usepackage{multirow}
\usepackage{float}
\setlist[enumerate]{leftmargin=*, itemsep=0.25\baselineskip}
\usepackage{xcolor}

\newcolumntype{Y}{>{\raggedleft\arraybackslash}X}

\begin{document} 

\defcitealias{margalef2024agn}{MB26}

   \title{Towards a consistent framework of determining active galactic nucleus contribution fraction and host galaxy properties}

   \author{Mi Chen \inst{1}\thanks{\email{chenmi@astro.rug.nl}},
          Lingyu Wang \inst{1, 2},
          Berta Margalef-Bentabol \inst{2}, 
          Katarzyna Małek \inst{3, 4},
          Brivael Laloux \inst{5}, 
          \and
          Antonio La Marca \inst{1,2, 6}.
          }

   \institute{Kapteyn Astronomical Institute, University of Groningen,
              Postbus 800, 9700 AV Groningen, The Netherlands
         \and
             SRON Netherlands Institute for Space Research, Landleven 12, 9747 AD Groningen, The Netherlands
         \and
             National Centre for Nuclear Research, Pasteura 7, 02-093, Warsaw, Poland
        \and 
            Aix-Marseille Université, CNRS, CNES, LAM, Marseille, France             
        \and
            INAF - Osservatorio Astronomico di Capodimonte Salita Moiariello 16, 80131, Napoli, Italy
        \and
            European Space Agency/ESTEC, Keplerlaan 1, 2201 AZ Noordwijk, The Netherlands
             }

   \date{Received -; accepted -}

\abstract{Decomposing active galactic nuclei (AGN) contributions from their host galaxies is essential for  identifying AGN-dominated galaxies and accurately deriving key physical properties of both supermassive black hole (SMBHs) and their host galaxies, such as black hole accretion rates, stellar masses, and star-formation rates. 
However, decomposing AGN contributions from multi-wavelength photometry remains challenging due to inherent parameter degeneracies in spectral energy distribution (SED) fitting.
We establish a unified framework for estimating AGN contribution fractions ($f_{\mathrm{AGN}}$) and host galaxy properties by combining complementary AGN diagnostics: SED decomposition from two independent SED-fitting codes (CIGALE and GRAHSP) and deep learning (DL) based imaging decomposition. We apply SED decomposition to galaxies in the COSMOS-Web field using multi-wavelength photometry (ultraviolet to far-infrared) and calculate the AGN-to-host galaxy flux ratio  ($f_{\mathrm{AGN}}^{\mathrm{SED}}$) in the JWST/NIRCam F150W filter. We then compare these SED-fitting derived AGN fractions with those obtained from image decomposition using DL ($f_{\mathrm{AGN}}^{\mathrm{DL}}$).  
Strong agreement is found for AGN-negligible galaxies across all methods: $\sim$82\% are identified as such by both SED codes ($f_{\mathrm{AGN}}^{\mathrm{SED}}<0.1$), and $\sim$87\% of these also show minimal AGN contribution in the DL model. This high concordance demonstrates that current methods reliably identify pure galaxy systems in most cases.
However, for potential AGN-dominant galaxies, large discrepancies emerge. Only 2.7\% of galaxies show consistent significant AGN identification between SED codes, and merely $\sim$7.6\% of these also exhibit significant AGN fraction in DL estimation. 
Forcing DL AGN fraction constraints into SED fitting generally works for SED-significant/imaging-negligible cases but fails in around 30\% of SED-negligible/imaging-significant cases.
In summary, our results highlight the degeneracies inherent in current SED-fitting methods based on empirical observations or theoretical templates and demonstrate the power of incorporating  independent morphological information to break these degeneracies. We  present a more  robust framework combining morphological information from high-fidelity imaging from surveys such as the James Webb Space Telescope (JWST) and \textit{Euclid}, enabling improved AGN–host galaxy decomposition and more reliable measurements of both AGN and galaxy properties.} 

   \authorrunning{M. Chen et al.}
   \titlerunning{Towards a consistent framework of determining AGN contribution fraction}

   \maketitle

\section{Introduction}

Active Galactic Nuclei (AGN) are central regions in galaxies powered by accretion of material onto supermassive black holes \citep[SMBH;][]{richstone1998supermassive}. Many observed correlations between SMBHs and their host galaxies, such as black hole mass vs. stellar mass ($M_{\mathrm{BH}}$-$M_\star$) relation, BH mass vs. velocity dispersion ($M_{\mathrm{BH}}$-$\sigma$) relation, and $M_{\mathrm{BH}}$ vs. dynamical mass ($M_\mathrm{dyn}$) relation \citep{tremaine2002slope,reines2015relations, ding2020mass, menci2023outflows}, indicate causal connections or co-evolution. For example, AGN can significantly influence galaxy evolution through both radiative and mechanical processes \citep{pierce2019agn, harrison2024observational, dyda2025time}. These various feedback mechanisms lead to interconnected and multi-phased outcomes. For example, interactions of AGN-driven outflows and jets with host galaxies can regulate their star formation and deplete their cold gas reservoirs \citep{karouzos2014tale, guo2022cold, bessiere2024qsofeed}.

The diverse and complex physical processes in AGN allow them to emit radiation across the electromagnetic spectrum, with each wavelength band tracing different physical origins and mechanisms. Consequently, a wide range of identification methods has been developed, spanning from X-ray to radio wavelengths, e.g. mid-infrared (MIR) colour selection, optical emission-line diagnostics, and X-ray or radio observational properties (for a review, see \citealt{padovani2017active}). However, as each selection method traces distinct physical components of AGN, these methods are inherently subject to strong selection biases and cannot provide a complete census of the AGN population. For instance, (soft) X-ray selection is known to be biased against obscured sources, particularly Compton-thick AGN \citep{hickox2018obscured}. MIR broadband selection, relying on the generally redder colours of AGN, often suffers from contamination by star-forming galaxies and high-redshift massive galaxies \citep{hainline2016mid}. Optical spectroscopic selection, on the other hand, becomes more unreliable in galaxies with intense star formation \citep{lyu2022agn}. As a result, AGN samples identified by different methods exhibit only partial overlap, and no single method can provide a complete and unbiased AGN census.

To mitigate selection biases, one of the state-of-the-art approaches is to decompose the spectral energy distribution (SED) into contributions from AGN and its host galaxy (for a review, see \citealt{conroy2013modeling}). SED fitting employs libraries of model spectra, including stellar populations, AGN radiation, dust re-emission, and nebular emission, and fits them to multi-wavelength photometric data. When covering a broad wavelength range, e.g. from the far-ultraviolet (FUV) to the far-infrared (FIR), SED fitting can account for multiple physical processes simultaneously, making it a potentially more comprehensive and versatile tool for AGN studies. Following SED decomposition, AGN can be identified by applying a manually defined selection criterion to the derived AGN contribution fraction $f_{\mathrm{AGN}}$ \citep{mountrichas2021x}. 
Conventionally, $f_{\mathrm{AGN}}$ is defined as the ratio of the AGN luminosity to the combined luminosity from both AGN and dust emission.
Alternatively, the difference in goodness-of-fit between models with and without an AGN component also provides a method for identifying AGN \citep{pouliasis2020obscured}.

However, the robustness of SED decomposition remains debated. In SED fitting, the number of model parameters often exceeds the available photometric constraints, resulting in increased risk of parameter degeneracy. A well-known example is the age--dust--metallicity degeneracy, which becomes particularly severe when FIR photometry is unavailable \citep{tacchella2022stellar, papovich2001stellar}. Similar degeneracies can also occur between AGN and host galaxy emission \citep{burke2022dwarf, martinez2024agnfitter}, leading to inconsistencies in the inferred $f_{\mathrm{AGN}}$ among different fitting codes and initial parameter settings.  For example, \cite{pacifici2023art} applied 14 popular SED-fitting codes to the CANDELS samples at $z\sim1$ and $z\sim3$, finding significant degeneracies between AGN and star-formation contribution. The issue of large uncertainties in the predicted AGN fractions due to different AGN torus models is also demonstrated in \cite{papadopoulos2025comparative}.

To disentangle degeneracies between AGN and host galaxy emission in SED fitting, imaging surveys can provide valuable morphological information. Beyond the local universe, AGN typically appear as unresolved point sources (whose dusty torus spans $\lesssim 10$ pc), while their host galaxies are typically spatially extended on scales of several kpc to tens of kpc \citep{fathi2010scalelength, cackett2021reverberation}. This size distinction enables image decomposition via surface brightness fitting, using a point-spread function (PSF) for the AGN component and (typically) Sérsic profiles for the host galaxy \citep[e.g.][]{li2021sizes, toba2022erosita, zhuang2024active}. The decomposed fluxes yield a wavelength-dependent AGN contribution fraction ($f_{\mathrm{AGN, \lambda}}$), which can then be used for more refined SED fitting. 
For instance, \cite{zhuang2024active} performed a multi-wavelength surface brightness fitting for 143 X-ray-selected AGN and fitted host galaxy SEDs using the decomposed photometry. Another possibility is to jointly fit AGN and host-galaxy SED components using priors derived from image decomposition.
\cite{yu2024new} applied decomposition-based flux priors in Bayesian SED fitting to 24 type-I AGN host galaxies using HST/WFC3 imaging. 
However, the limited spatial resolution and small survey areas prior to the launch of the James Webb Space Telescope (JWST) and the \textit{Euclid} space telescope restricted statistical analyses on large samples at higher redshifts.

Large survey areas with high-fidelity imaging data have only recently become available through JWST and \textit{Euclid}. The vast data volume poses computational challenges for traditional surface brightness fitting methods \citep{chen2024galmoss}. Moreover, these methods can fail when applied to galaxies with complex morphologies or when a simple Sérsic profile cannot adequately describe the structure \citep{ribeiro2016size, ghosh2023morphological}. To address these limitations, deep learning (DL) models have been developed to directly decompose observed galaxy images into AGN and host galaxy components. For example, \citet[][hereafter \citetalias{margalef2024agn}]{margalef2024agn} trained a DL model using PSF-injected mock JWST images to derive $f_{\mathrm{AGN, \lambda}}$ from 25,596 galaxy images in the JWST/NIRCam F150W band. Subsequently, the same approach was fine-tuned and applied to the \textit{Euclid} Deep Fields (EDFs), yielding measurements for 624,153 galaxies \citep{margalef2025euclid}. 
However, the consistency between image-based and SED-based AGN contribution estimates remains largely unexplored for large galaxy samples. If systematic discrepancies exist, developing mitigation strategies will be essential.

Motivated by these considerations, we aim to develop a consistent framework of determining AGN contribution fraction and host galaxy properties using $\sim15,000$ galaxies from the COSMOS field.  Because this work measures $f_{\mathrm{AGN, \lambda}}$ exclusively in the JWST/NIRCam F150W band, we hereafter denote it simply as $f_{\mathrm{AGN}}$.
First, we compare AGN contribution fraction inferred from two independent SED-fitting codes ($f_{\mathrm{AGN}}^{\mathrm{SED}}$): Code Investigating GALaxy Emission \citep[CIGALE;][]{burgarella2025cigale, noll2009analysis,boquien2019cigale} and Genuine Retrieval of the AGN Host Stellar Population \citep[GRAHSP;][]{buchner2024genuine}. As CIGALE models the SED analytically while GRAHSP adopts an empirical approach, their combination helps mitigate systematic biases in estimating $f_{\mathrm{AGN}}^{\mathrm{SED}}$ and provides a more robust decomposition from the SED-fitting perspective. Next, we compare with the image decomposition results from AGN-to-host flux ratios ($f_{\mathrm{AGN}}^{\mathrm{DL}})$ reported by \citetalias{margalef2024agn}, in which a DL-based model decomposes the AGN component in JWST images. Finally, we discuss and present a workflow for adopting the preferred AGN fraction in both concordant and discrepant cases, along with the corresponding host galaxy properties.

This paper is structured as follows. Section~\ref{sec: data} introduces the photometric catalogues and sample selection used in this work, along with a brief summary of the DL-based imaging decomposition from \citetalias{margalef2024agn}. Section~\ref{sec: Methods} describes the two SED-fitting codes and defines the AGN contribution fraction, including its computation.
Section~\ref{sec: results} presents a comparison between AGN fractions inferred from the two SED-fitting methods and the DL-based imaging results. Section~\ref{sec: discussion} further investigates subsamples showing significant discrepancies between the methods. Finally, our conclusions and practical guidelines are summarised in Sect.~\ref{sec: conclusions} Throughout the paper, we adopt the AB magnitude system unless otherwise stated.

\section{Data}
\label{sec: data}

The galaxy sample used in this study is drawn from the Cosmic Evolution Survey (COSMOS; \citealt{scoville2007cosmic}) field, one of the most extensively observed extragalactic fields. 
COSMOS covers $\sim$2~$\mathrm{deg}^2$ with extensive spectroscopic and imaging data.
In this section, we first describe how we construct the photometric catalogue used for SED fitting by combining existing datasets (Sect.~\ref{subsec: catalogue}). We then outline the sample selection criteria in Sect.~\ref{subsec: sample selection}. In Sect.~\ref{subsec: MB26}, we describe the \citetalias{margalef2024agn} catalogue, which provides $f_{\mathrm{AGN}}^{\mathrm{DL}}$ estimates based on COSMOS-Web imaging.

\subsection{The multi-wavelength datasets in COSMOS}
\label{subsec: catalogue}

The primary catalogue used in this study is COSMOS2020 \citep{weaver2022cosmos2020}, which provides homogeneous multi-wavelength photometry from the ultraviolet (UV) to the MIR. 
It covers a wavelength range of $0.1$--$10\,\mu\mathrm{m}$ and contains approximately 1.7 million sources. The FUV and the Near-ultraviolet (NUV) data are obtained from the Galaxy Evolution Explorer (GALEX) satellite \citep{morrissey2007calibration}. 
The optical bands include U-band data from the Canada--France--Hawaii Telescope (CFHT) Large Area U-band Deep Survey (CLAUDS; \citealt{sawicki2019cfht}), grizy-band data from the Hyper Suprime-Cam Subaru Strategic Program (HSC-SSP) PDR2 \citep{aihara2019second}, and seven broad bands, together with 12 medium bands and two narrow bands from Subaru/Suprime-Cam \citep{taniguchi2007cosmic, taniguchi2015subaru}. 
The near-infrared (NIR) dataset includes YJHK$_\mathrm{s}$-band and narrow-band NB118 data from the fourth data release (DR4) of the UltraVISTA survey \citep{mccracken2012ultravista}, as well as Spitzer/Infrared Array Camera (IRAC; \citealt{werner2004spitzer}) imaging in channels 1-4. In total, 44 photometric bands are included in COSMOS2020. Alongside multi-wavelength photometry, derived physical parameters such as stellar mass, star-formation rate (SFR), and photometric redshift (photo-$z$) are also provided. However, the seven broad bands from Subaru/Suprime-Cam are not used in this study because HSC-SSP provides deeper observations in these bands.

We also utilise the latest infrared observations from the COSMOS-Web \citep[][PIs: Kartaltepe \& Casey, ID=1727]{casey2023cosmos} survey to better constrain SEDs in the NIR. As a 255-hour JWST Treasury Program, COSMOS-Web provides $\sim$0.6~$\mathrm{deg}^2$ contiguous Near Infrared Camera (NIRCam) imaging in four bands (F115W, F150W, F277W, F444W). In addition,  $\sim$0.2~$\mathrm{deg}^2$ of non-contiguous area is observed with Mid-Infrared Instrument (MIRI) in the F770W filter. We cross-match and add these five bands from the COSMOS2025 catalogue \citep{shuntov2025cosmos2025}, which shows consistent photometry with COSMOS2020 in all overlapping bands.

We further extend to the FIR regime by cross-matching with the catalogue from \cite{wang2024probabilistic}, which provides deblended FIR and submillimetre point source fluxes. The deblending is performed sequentially with XID+ \citep{hurley2017help}, a Bayesian probabilistic framework. Initial priors are derived from COSMOS2020 and radio catalogues to deblend the {\it Spitzer}/MIPS 24 µm data.
The deblended 24 µm fluxes then update the priors for deblending the {\it Herschel}/PACS 100 and 160 µm maps, which in turn provide priors for the {\it Herschel}/SPIRE 250, 350, and 500 µm observations.
The FIR extension provides crucial information on dust emission, helping to mitigate the degeneracy between AGN and host galaxy components in SED fitting. Additionally, we compiled the X-ray catalogue from \cite{la2026major}, which derives from the Chandra COSMOS Legacy survey \citep{civano2016chandra, marchesi2016chandra}.

In total, our catalogue includes 46 photometric bands spanning from the X-ray to the FIR, covering approximately 960,000 galaxies. A summary and brief description of each band, particularly those newly added beyond COSMOS2020, are provided in Table~\ref{tab: catalogue}.

\subsection{Sample selection}
\label{subsec: sample selection}
To ensure a consistent sample when integrating the $f_{\mathrm{AGN}}^{\mathrm{DL}}$ measurements, we restrict our sample to the COSMOS-Web region used in the \citetalias{margalef2024agn} study. We utilise the Farmer catalogue for photometric parameters. We adopt photometric redshifts (photo-$z$) from the LePhare code \citep{arnouts2002measuring, ilbert2006accurate}, and replace them with spectroscopic redshifts (spec-$z$) when available. Stellar masses are taken from \cite{la2026major}, whose estimates incorporate AGN templates using CIGALE, and are consistent with the LePhare stellar mass reported in the COSMOS2020 catalogue.

We select a clean galaxy sample following the criteria recommended by the COSMOS team: 
\begin{align}
&\text{FLAG\_COMBINED} = 0, \label{eq:flag_combined} \\
&\text{lp\_type} = 0\ (\text{galaxy})\ \text{or}\ 2\ (\text{X-ray source}),
\end{align}
where \text{FLAG\_COMBINED} selects objects detected in HSC, UltraVISTA, and {\it Spitzer}/IRAC images while excluding those affected by bright stars or significant image artefacts. The flag \text{lp\_value} distinguishes between stars and galaxies based on photometric redshifts and physical parameters. 
After applying the quality cuts, we construct a mass-complete sample based on two criteria. The first is the $70\%$ mass completeness threshold derived by \cite{weaver2022cosmos2020} based on the $K_S$ limit:
\begin{align}
M_{\mathrm{lim}}(z) &= -3.55 \times 10^8 (1 + z) + 2.70 \times 10^8 (1 + z)^2. \label{eq: m_com}
\end{align}
The second criterion requires galaxies to have stellar mass $M_*> 10^9 M_\odot$, reflecting the limit for the simulated galaxy sample used to train the DL model in \citetalias{margalef2024agn} (see \cite{margalef2024agn} for details). For each galaxy, we adopt the higher of these two limits. 

The redshift range is restricted to $0.5 \leq z \leq 3$, consistent with \citetalias{margalef2024agn}. In addition, since \citetalias{margalef2024agn} uses galaxy imaging from \cite{zhuang2024active}, the sample is further limited to galaxies located within the same 0.28 $\mathrm{deg}^2$ field. The resulting final sample comprises 25,596 galaxies.

\subsection{The imaging-derived AGN contribution fraction}
\label{subsec: MB26}

\cite{margalef2024agn} utilises a DL convolutional neural network (CNN) model fine-tuned from Zoobot \citep{walmsley2023zoobot} to estimate the fractional contribution of the central point source in the observed galaxy image. 

The DL model is trained on mock JWST images, combining simulated galaxies from IllustrisTNG \citep{pillepich2018simulating} with real sky backgrounds from JWST/NIRCam F150W. To mimic AGN host galaxies, corresponding JWST PSF models are injected. By varying the luminosity of the injected PSF, galaxies with different levels of AGN contribution are included in the training sample. 
The model demonstrated excellent performance on the test subset. The mean difference between injected $f_{\mathrm{AGN}}$ and predicted $f_{\mathrm{AGN}}$ is -0.0018 with an RMSE = 0.013 overall. 
The model also shows good agreement with image decomposition results from \texttt{GALFIT} \citep{peng2010detailed}, which combines a PSF function and a Sérsic profile. 

After training, the model is applied to real JWST/NIRCam F150W images to predict the AGN contribution fraction, which we denote as $f_{\mathrm{AGN}}^{\mathrm{DL}}$ hereafter. These results are published in \citetalias{margalef2024agn}.

\section{Methods}
\label{sec: Methods}

The extensive multi-wavelength photometric coverage available in COSMOS provides a unique opportunity to constrain the relative flux contributions of stellar, dust, and AGN components to the observed SEDs of galaxies. 
The AGN contribution fraction ($f_{\mathrm{AGN}}^{\mathrm{SED}}$) is defined as the ratio of the integrated AGN model flux to the total model flux over a given wavelength range. 
The derived $f_{\mathrm{AGN}}^{\mathrm{SED}}$ reflects the relative flux contribution of the unresolved AGN (point source) and the extended host galaxy, and is thus expected to be consistent with the galaxy morphology inferred from high-resolution imaging within the same wavelength range. 
For example, a point source-dominated morphology should correspond to a large $f_{\mathrm{AGN}}^{\mathrm{SED}}$. 
High spatial resolution is essential for separating AGN emission from the host galaxy. Previously, such AGN-host decomposition was feasible primarily with the \textit{Hubble} Space Telescope (HST) for high-redshift galaxies. 
With the advent of JWST and \textit{Euclid}, wide-field imaging surveys now achieve substantially larger sky coverage while maintaining spatial resolution comparable to (or better than) that of HST. This advance enables AGN--host decomposition for statistically large galaxy samples, as demonstrated by the \citetalias{margalef2024agn} catalogue. 

Having both $f_{\mathrm{AGN}}^{\mathrm{SED}}$ and $f_{\mathrm{AGN}}^{\mathrm{DL}}$ enables a direct comparison between these independent measurements, allowing us to evaluate the consistency and robustness of AGN fraction determinations from SED decomposition and image decomposition techniques. To derive $f_{\mathrm{AGN}}^{\mathrm{SED}}$, we model the SEDs of our sample galaxies using two different codes: CIGALE \citep{boquien2019cigale} and GRAHSP \citep{buchner2024genuine}. For the galaxy components, both codes rely on the same template libraries within the CIGALE framework. The main difference lies in the AGN modelling. In CIGALE, AGN emission is represented by a single AGN module based on physically motivated templates. In comparison, GRAHSP decomposes AGN emission into three empirically derived sub-components—continuum, emission lines, and torus emission—allowing for a more flexible description of the AGN contribution. As a result, the shapes of the AGN templates differ between the two libraries. As an illustration, for a given FIR normalisation, the AGN templates can exhibit different spectral extensions into the shorter-wavelength regime (see Fig.~\ref{fig: agn library}  and Appendix~\ref{app:library_diff}).
In the following section, we describe the detailed configuration of each code.

\begin{figure*}[htp]
    \sidecaption
    \centering
    \includegraphics[width=0.7\textwidth]{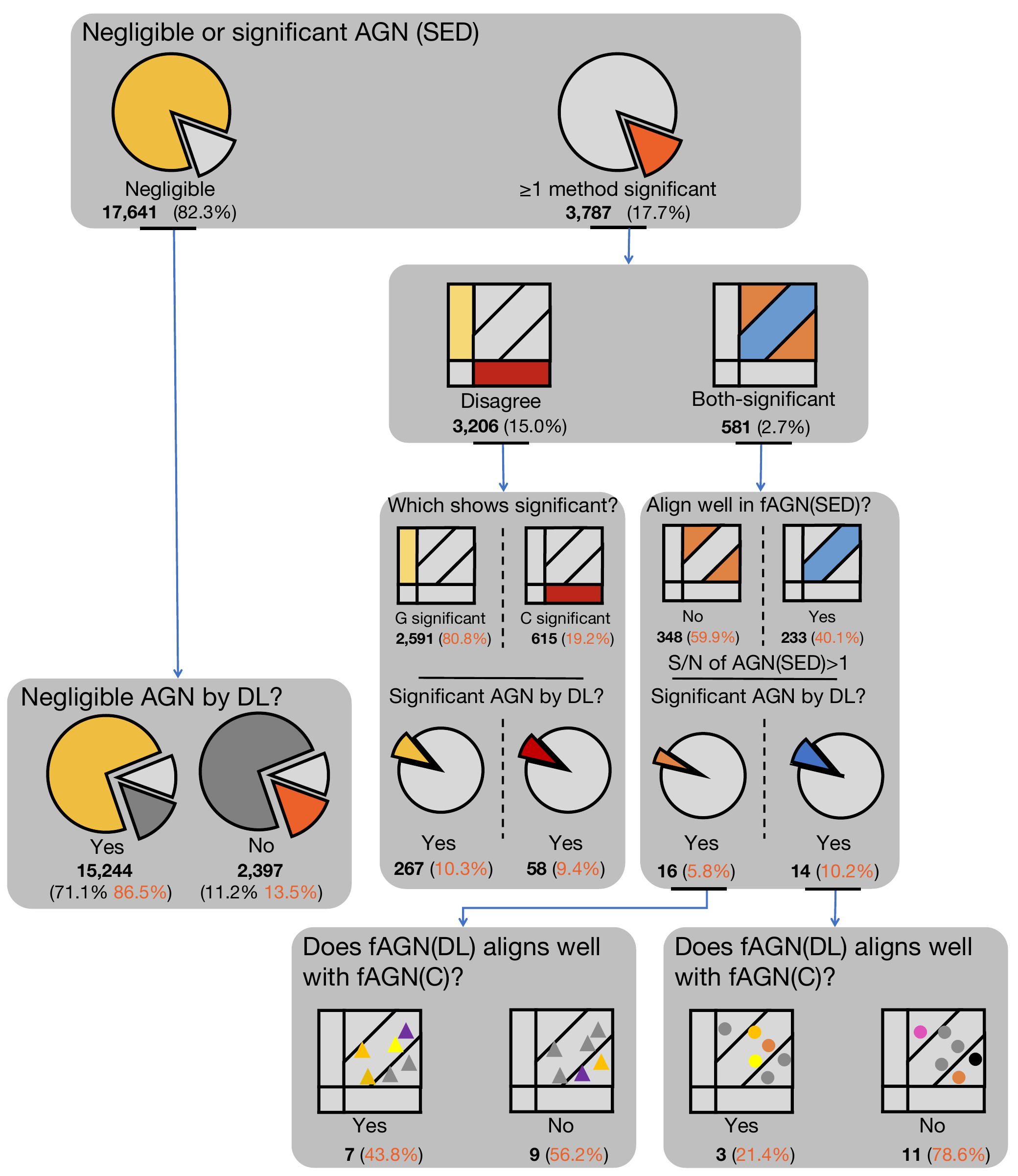}
    \caption{Processing steps and summary statistics in our comparison framework. We first compare CIGALE and GRAHSP to see if they agree on a negligible AGN contribution fraction ($f_{\mathrm{AGN}}$) in the top box. Next, SED AGN-negligible galaxies are directly compared with the DL results, as shown in the left-hand flowchart. On the right, we analyse galaxies for which at least one of the SED fitting methods identifies a significant AGN contribution. Within this flowchart, we begin by comparing the results from two SED fitting methods (see Fig.~\ref{fig: comparison_sed}). We then incorporate the DL results (see Fig.~\ref{fig: comparison} and Fig.~\ref{fig: comparison_ABC}). We note that an additional S/N selection ($\mathrm{S/N}_{\mathrm{AGN(SED)}} > 1$) is applied before the comparison with DL, specifically for the branch starting from the “Align well in $f_{\mathrm{AGN(SED)}}$?” step in the right-hand flowchart. Each grey box represents a step and displays the corresponding fraction of galaxies. Black percentages are calculated relative to the parent sample, with small-number cases shown as counts, while orange percentages indicate conditional fractions from the previous step. Overall, we find good agreement for AGN-negligible galaxies, whereas significant discrepancies emerge in the AGN-significant regime.}
    \label{fig:processing}
\end{figure*}

\subsection{SED modelling: CIGALE}

CIGALE has been widely adopted for SED-fitting studies. Since its major Python rewrite \citep{boquien2019cigale}, the code has incorporated additional components, including X-ray emission \citep{yang2020x, yang2022fitting}. In this work, we use CIGALE version 2025.0\footnote{\url{https://cigale.lam.fr/2025/01/17/version-2025-0/}}.
CIGALE operates under the energy balance assumption: the energy absorbed by dust in the UV--NIR is re-emitted self-consistently at longer wavelengths in the MIR and FIR. In our configuration, the stellar component is generated from the single stellar population templates of \cite{bruzualStellarPopulationSynthesis2003}. After being reddened by dust following \cite{charlotSimpleModelAbsorption2000}, the absorptions in the UV--NIR are re-emitted self-consistently at longer wavelengths in the MIR and FIR using the model of \cite{draineAndromedasDust2014}.

In addition to the stellar and dust components, CIGALE allows for the inclusion of an AGN dust emission component. Two AGN templates are implemented: the smooth torus model \texttt{Fritz} \citep{fritz2006revisiting} and the clumpy torus model \texttt{SKIRTOR} \citep{stalevski2016dust}. The former adopts a flared-disk torus geometry with silicate and graphite grains of various sizes, and computes the AGN SED via two-dimensional radiative transfer calculations. In contrast, \texttt{SKIRTOR} models a clumpy medium embedded within a smooth dust component, using three-dimensional radiative transfer. This two-phase geometry reproduces the observed MIR anisotropy more realistically, making \texttt{SKIRTOR} more consistent with current observations \citep{yang2020x, cruz2023modeling}. We perform SED-fitting runs using both templates while keeping all other modules fixed to assess potential systematic differences. Additionally, the X-ray module is utilised for AGN and galaxy X-ray emission \citep{yang2020x, yang2022fitting}.

CIGALE decomposes SEDs into four components: attenuated stellar emission, dust emission, nebular emission, and AGN emission. The best-fitting physical parameters corresponding to the minimum $\chi^2$ are recorded. In addition, for key quantities such as stellar mass, SFR, and dust attenuation $\mathrm{A_V}$, we enable CIGALE to report Bayesian estimates and uncertainties derived from the likelihood-weighted distribution of models, corresponding approximately to the peak of the posterior probability distribution.

The \texttt{AGN fraction} parameter in CIGALE configuration (as shown in Tab. \ref{tab:cigale_parameters}) is defined over a user-specified rest-frame wavelength interval. To enable a direct comparison with the CNN-predicted $f_{\mathrm{AGN}}^{\mathrm{DL}}$ from \citetalias{margalef2024agn}, derived from observed-frame JWST/NIRCam F150W images, we add a parameter to the CIGALE redshift module that computes an observed-frame $f_{\mathrm{AGN}}$. This parameter is calculated by integrating the decomposed AGN and total SEDs through the JWST/NIRCam F150W bandpass:
\begin{equation}
\label{eq:fagn_cigale}
f_{\mathrm{AGN}}^{\mathrm{C}} = 
\frac{\int T_{\mathrm{F150W}}(\lambda)\, F_{\mathrm{AGN}}(\lambda)\, d\lambda}
{\int T_{\mathrm{F150W}}(\lambda)\, F_{\mathrm{tot}}(\lambda)\, d\lambda},
\end{equation}
where $T_{\mathrm{F150W}}$ is the JWST/NIRCam F150W transmission curve\footnote{\url{https://jwst-docs.stsci.edu/jwst-near-infrared-camera/nircam-instrumentation/nircam-filters}}, $F_{\mathrm{AGN}}$ is the AGN SED, and $F_{\mathrm{tot}}$ is the total SED including stellar, dust, and AGN emission. 
We find good agreement between \texttt{SKIRTOR} and \texttt{Fritz}, with 85\% of galaxies showing a difference $\Delta f_{\mathrm{AGN}}^{\mathrm{C}} < 0.1$. Given its more physically motivated two-phase torus geometry, we adopt the \texttt{SKIRTOR}-based $f_{\mathrm{AGN}}^{\mathrm{C}}$ in subsequent analyses. Our full initial parameter grid is listed in Table~\ref{tab:cigale_parameters}.

\subsection{SED modelling: GRAHSP}

Built upon the CIGALE framework, GRAHSP replaces the analytical models of AGN and dust attenuation with empirical templates, incorporating power-law continuum emission from the accretion disk, emission lines, and infrared torus emission. The physical diversity of AGN structures and emission mechanisms (often referred to as the “AGN zoo”) remains a subject of debate, and no single analytical model can fully reproduce the photometric properties of AGN--host systems. The empirical approach adopted in GRAHSP offers greater flexibility in capturing the observed diversity of AGN spectral features, enabling a more accurate description of AGN-dominated SEDs. When evaluated against CHIMERA \citep{johannes_buchner_2023_8431646} as a benchmark, GRAHSP achieves more accurate results than various CIGALE configurations explored.

For the galaxy component, we adopt the same configuration as used in CIGALE, including a delayed star-formation history with an optional exponential burst, the single stellar population templates from \cite{bruzualStellarPopulationSynthesis2003}, and the dust emission models from \cite{draineAndromedasDust2014}. For the AGN component, we activate the empirical modules: \texttt{activatepl} for the big blue bump continuum, \texttt{activatelines} for both narrow and broad emission lines, and \texttt{activategtorus} for the re-emission of dust-processed UV light by the torus. Both the AGN and host components are subject to dust attenuation, which is treated separately for each component through \texttt{E(B-V)} (host galaxy dust) and \texttt{E(B-V)-AGN} (nuclear dust) within the \texttt{biattenuation} module, allowing for distinct absorption processes between host galaxy light and AGN light. We do not impose a prior on the \texttt{AGN fraction} parameter in GRAHSP, following the recommendation of the GRAHSP developers. Instead, the AGN contribution is constrained through scaling limits on the AGN luminosity normalisation. Since GRAHSP does not explicitly include an X-ray emission component, we exclude X-ray bands from the SED fitting.
The initial fitting parameters for GRAHSP are summarised in Table~\ref{tab:grahsp_parameters}. 

GRAHSP employs a Bayesian framework with nested sampling, which efficiently explores multi-modal posterior distributions and ensures robust convergence. Unlike CIGALE, which evaluates galaxy and AGN parameter grids jointly, GRAHSP explores the two parameter spaces separately. When the sampler explores the AGN parameter space, the galaxy component is replaced by a simplified mock model, and vice versa.
After both components have been computed, the two SEDs are combined, with their relative amplitudes optimised to minimise $\chi^2$ for each sampling step. In this way, the AGN-to-galaxy contribution in GRAHSP is inferred naturally from the data, rather than being imposed by predefined scaling parameters as in grid-based frameworks such as CIGALE. In the output, GRAHSP can directly output the bandpass-integrated AGN flux with its uncertainty, including contributions from the accretion disk, torus emission, and broad/narrow emission lines. The observed-frame AGN fraction in the JWST/NIRCam F150W band can therefore be computed straightforwardly as:
\begin{equation}
\label{eq:fagn_grahsp}
f_{\mathrm{AGN}}^{\mathrm{G}} = 
\frac{F_{\mathrm{AGN(F150W)}}}
{F_{\mathrm{tot(F150W)}}},
\end{equation}
where $F_{\mathrm{AGN(F150W)}}$ represents the total AGN flux transmitted through the JWST/NIRCam F150W filter, and $F_{\mathrm{tot(F150W)}}$ denotes the corresponding total galaxy flux in the same band. The uncertainty in $f_{\mathrm{AGN}}^{\mathrm{G}}$ is computed using standard error propagation.

\begin{figure}
    \centering
    \includegraphics[width=0.495\textwidth]{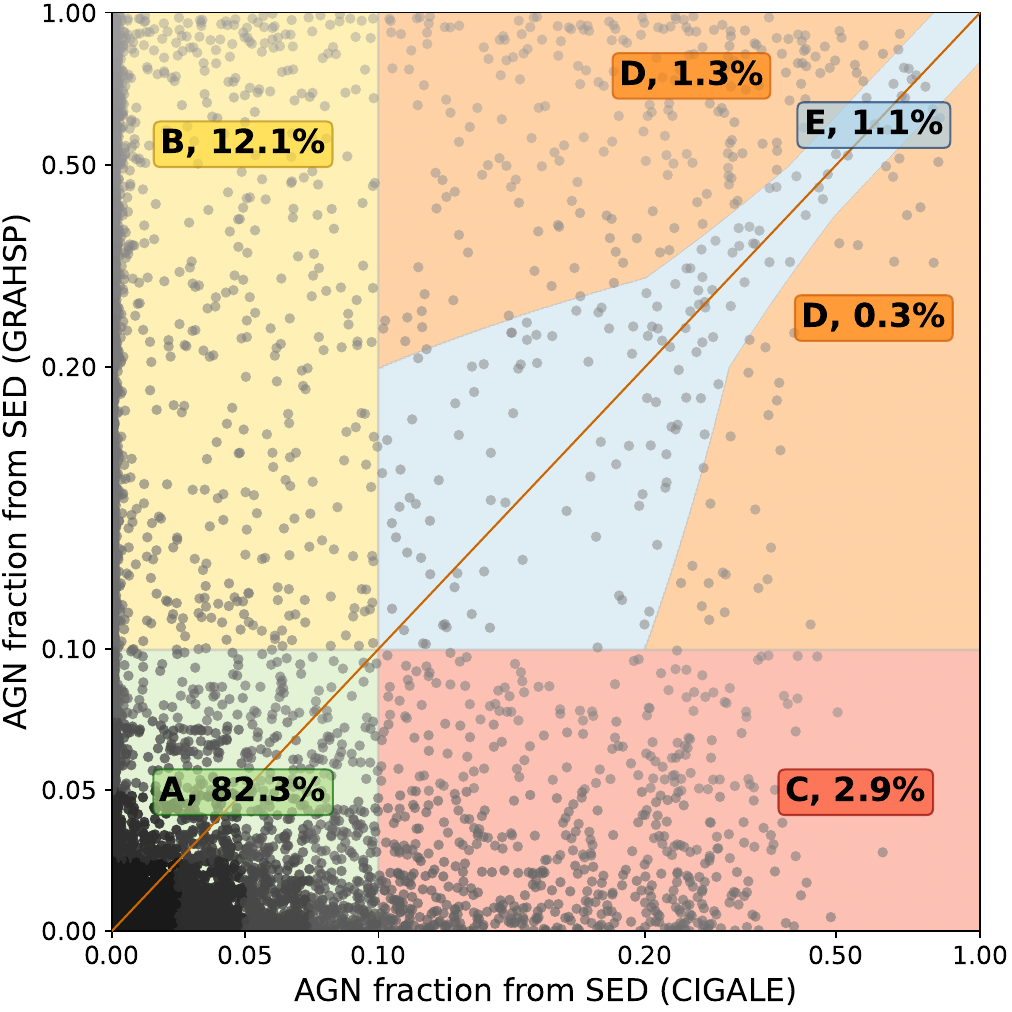}
    \caption{Comparison of $f_{\mathrm{AGN}}$ from CIGALE and GRAHSP. Grey dots represent the parent sample ($N = 21,428$), with point transparency scaled by local source density (number of sources per two-dimensional bin in the $f_{\mathrm{AGN}}$ plane). The orange diagonal line indicates perfect agreement (one-to-one relation). Coloured background regions delineate the sample classification: Region A (green) denotes both-negligible galaxies; Regions B+C (yellow and red) denote mixed-detection galaxies; Regions D+E (orange and blue) denote both-significant galaxies with scattered and consistent AGN fractions, respectively.}
    \label{fig: comparison_sed}
\end{figure}

\begin{figure*}
    \centering
    \includegraphics[width=0.995\textwidth]{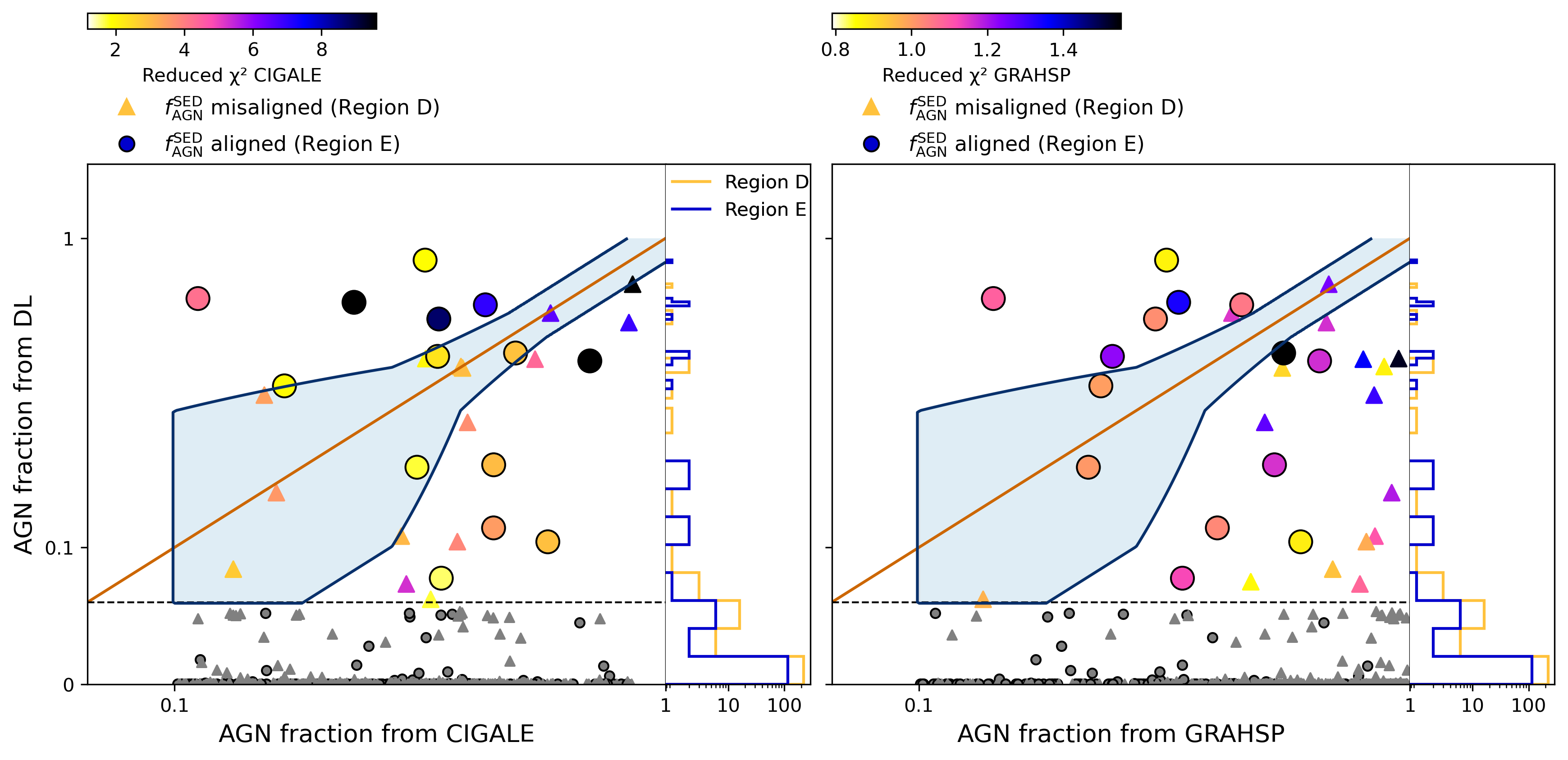}
    \caption{AGN contribution fractions derived from SED and DL are compared for galaxies in Regions D and E of Fig.~\ref{fig: comparison_sed}. The left panel compares $f_{\mathrm{AGN}}^{\mathrm{C}}$ with $f_{\mathrm{AGN}}^{\mathrm{DL}}$, while the right panel shows $f_{\mathrm{AGN}}^{\mathrm{G}}$ versus $f_{\mathrm{AGN}}^{\mathrm{DL}}$. The histograms show the 1D distribution of $f_{\mathrm{AGN}}^{\mathrm{DL}}$ for each samples. The orange diagonal indicates the one-to-one relation. The grey dashed horizontal line marks the $5\sigma$ threshold from \citetalias{margalef2024agn}. The blue shaded region marks where GRAHSP and CIGALE show reasonable agreement (see text for details). Triangles correspond to galaxies in Region D, while the circles correspond to those in Region E. Galaxies that are identified as AGN-significant by both SED and DL are colour-coded according to their normalised $\chi^2_\nu$, while grey points below the $5\sigma$ threshold indicate systems with negligible $f_{\mathrm{AGN}}^{\mathrm{DL}}$.}
    \label{fig: comparison}
\end{figure*}

\section{Results}
\label{sec: results}

In this section, we compare AGN fractions obtained from SED fitting and those predicted by the DL models in \citetalias{margalef2024agn}.
In principle, both methods should return consistent estimates, as they measure the relative flux contribution of the AGN and its host galaxy over the same wavelength range. However, as we show below, systematic discrepancies arise even between different SED-fitting codes, particularly at non-negligible AGN contribution fractions.  
The overall workflow and a summary of key number statistics are illustrated in Fig.~\ref{fig:processing}: starting from the parent sample selected through a series of SED-fitting quality cuts, we first compare AGN contribution fractions from CIGALE and GRAHSP ($f_{\mathrm{AGN}}^{\mathrm{C}}$ vs. $f_{\mathrm{AGN}}^{\mathrm{G}}$). After identifying a robust SED-based subsample, we incorporate the $f_{\mathrm{AGN}}^{\mathrm{DL}}$  values to assess their consistency with the SED-derived estimates.

\subsection{SED-fitting quality cuts}
To ensure reliable comparisons, we apply several SED-fitting quality cuts: 
\begin{flalign}
    \label{eq:sfr}  &\ \frac{1}{5} \le \frac{X_{\mathrm{best}}^{\mathrm{C}}}{X_{\mathrm{bayes}}^{\mathrm{C}}}  \le 5, &\\[6pt]  
    \label{eq:gussion} &\ \log \chi^{2}_{\nu, (\mathrm{C})} < 5\sigma \quad (\Delta z = 0.5), 
\end{flalign}
Eq. (\ref{eq:sfr}) is used for a best/Bayes ratio sanity check for parameters within a factor of five \citep{mountrichas2021galaxy}, where $X_{\mathrm{best}}^{\mathrm{C}}$ denotes the best-fit value, while $X_{\mathrm{bayes}}^{\mathrm{C}}$ is Bayesian posterior estimates derived by CIGALE. The parameter $X$ represents $\mathrm{SFR}$ and $\mathrm{M}_{\ast}$.
In Eq. (\ref{eq:gussion}), we examine the $\chi^2_\nu$ distribution to further assess the reliability of the fits. For CIGALE, within redshift bins of $\Delta z = 0.5$, we fit a Gaussian to each histogram of log $\chi^2_\nu$ values and discard sources exceeding the $5 \sigma$ limit (see Table~\ref{tab: chi2}). For GRAHSP, all sources have $\chi^2_\nu<4$, and no additional cuts are applied. 
After applying the quality cuts, the sample is reduced to 21,428 galaxies. We refer to this as our parent sample.

\begin{figure*}
    \centering
    \includegraphics[width=0.995\textwidth]{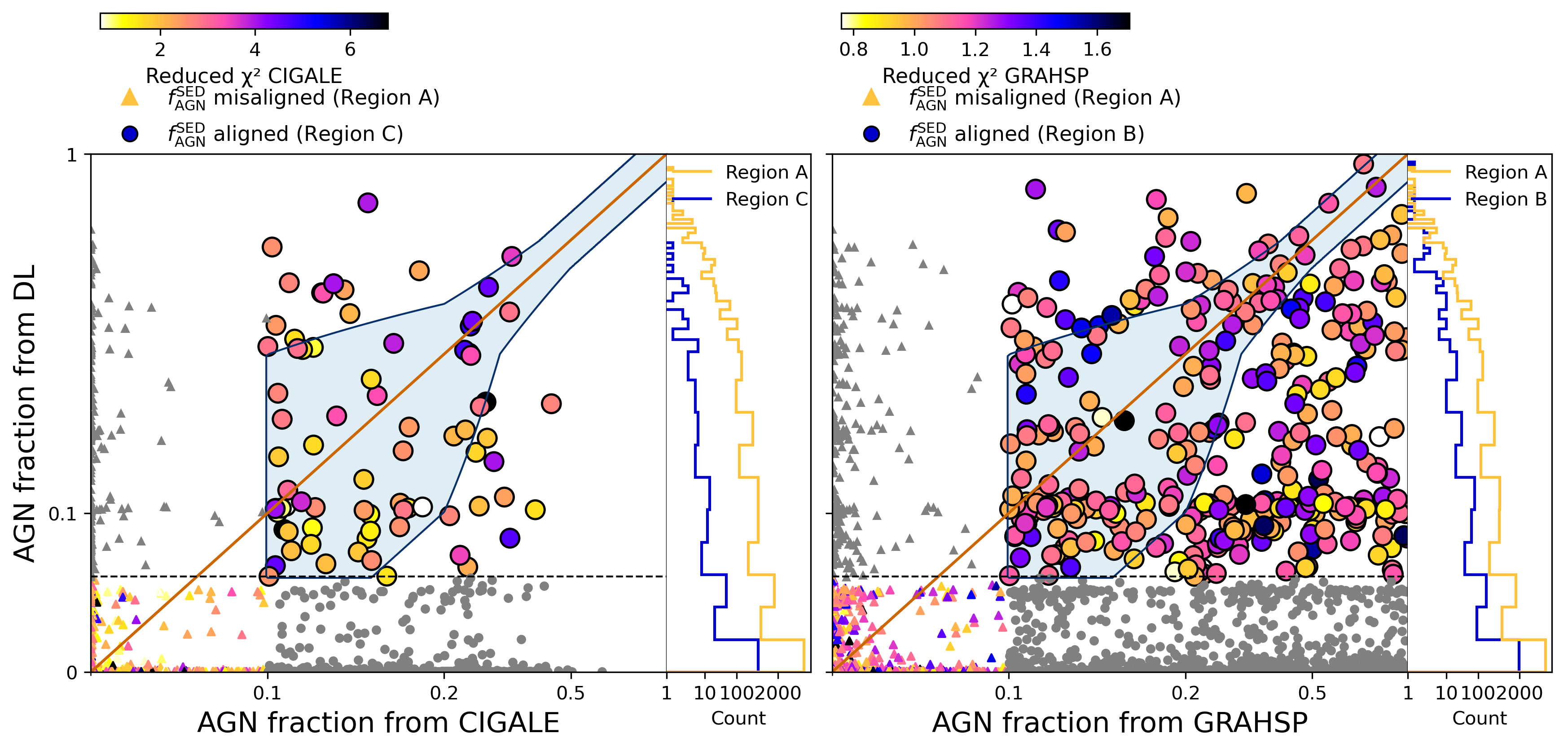}
    \caption{Following the same format as Fig.~\ref{fig: comparison}, AGN contribution fractions derived from SED and DL are compared for galaxies in Regions A, B, and C. Triangles represent galaxies in Region A, while circles correspond to galaxies in Region C (left panel) and Region B (right panel). Galaxies with consistent AGN classifications between SED and DL are colour-coded according to their normalised $\chi^2_\nu$, whereas those with inconsistent classifications are shown in grey.}
    \label{fig: comparison_ABC}
\end{figure*}

\subsection{Comparison between CIGALE and GRAHSP}

\begin{table}
\centering
\renewcommand{\arraystretch}{1.3} 
\caption{5$\sigma$ limits of the distribution of $\log \chi^2_\nu$ in CIGALE, in five redshift bins.}
\label{tab: chi2}
\begin{tabular}{c|ccccc}
\hline
$z$ range & 0.5--1.0 & 1.0--1.5 & 1.5--2.0 & 2.0--2.5 & 2.5--3.0 \\
\hline 
5$\sigma$ limit & 12.81 & 14.67 & 16.25 & 13.69 & 12.19 \\
\hline
\end{tabular}
\end{table}

We begin by comparing AGN contribution fractions obtained from the two SED-fitting methods, $f_{\mathrm{AGN}}^{\mathrm{C}}$ and $f_{\mathrm{AGN}}^{\mathrm{G}}$ (Fig.~\ref{fig: comparison_sed}). A nonlinear axis scaling is adopted to 
enhance the dynamic range at low $f_{\mathrm{AGN}}$, allowing differences between the two estimates to be more clearly visualised. Values below $f_{\mathrm{AGN}}=0.5$ occupy three-quarters of the axis range, while the remaining interval is compressed into the upper quarter. This scaling is used consistently in all subsequent figures involving AGN fraction comparisons.

First, we classify the 21,428 galaxies according to whether CIGALE and GRAHSP agree on the presence of a significant AGN component:
\begin{equation}
\label{eq:fagnsig}
f_{\mathrm{AGN}}^{\mathrm{SED}} > 0.1,
\end{equation}
which leads to three categories:
\begin{enumerate}
    \item Both-negligible (Region A): neither method identifies a significant AGN contribution fraction.
    \item Mixed-detection (Regions B and C): only one method identifies a significant AGN contribution fraction.
    \item Both-significant (Regions D and E): both methods identify a significant AGN contribution fraction.
\end{enumerate}
These categories contain 17,641 ($\sim82.3\%$), 3,206 ($\sim15.0\%$), and 581 ($\sim2.7\%$) galaxies, respectively.

\begin{figure}
    \centering
    \includegraphics[width=0.495\textwidth]{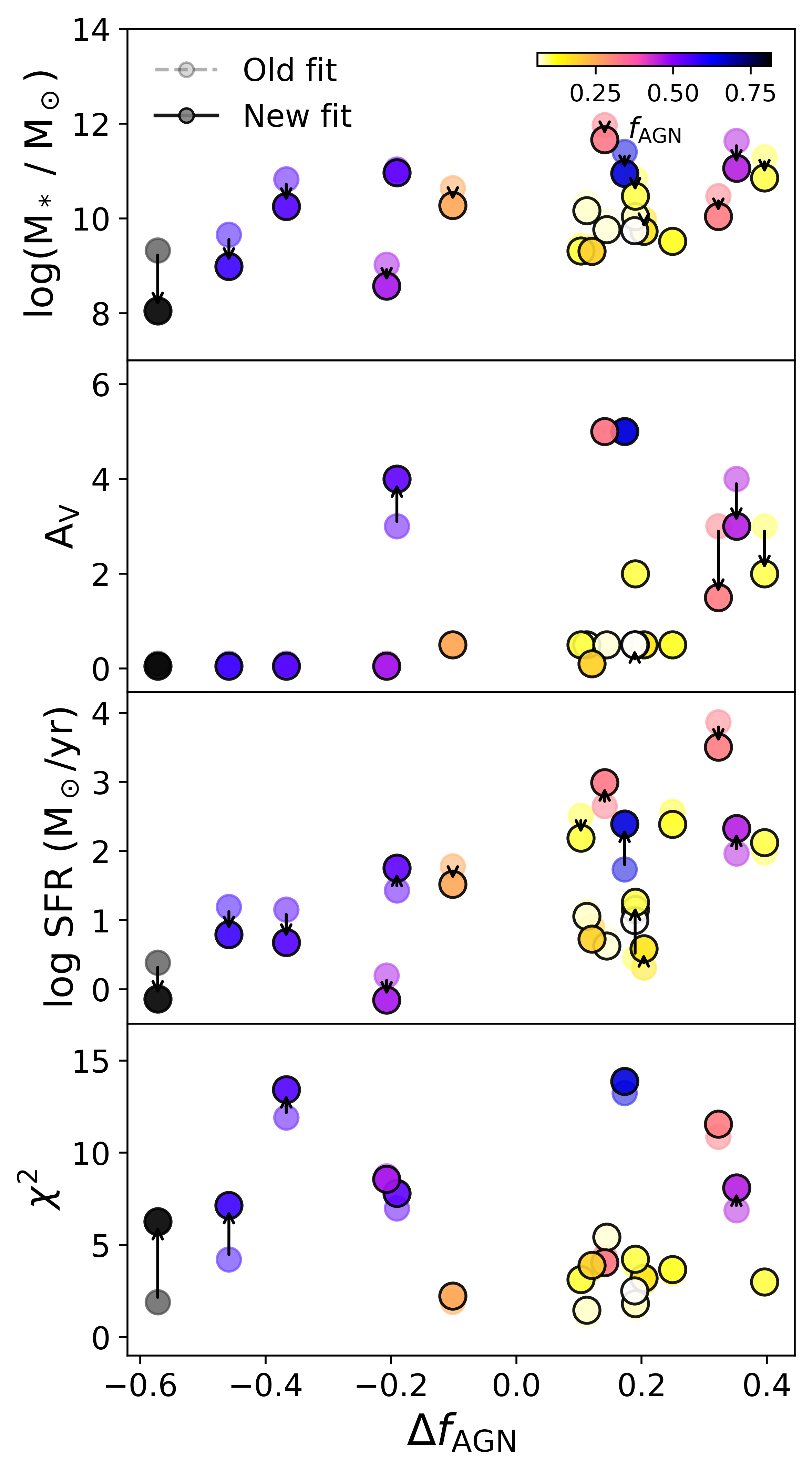}
    \caption{Relationship between the change in AGN fraction $\Delta f_{\mathrm{AGN}}$ (difference between old/new CIGALE fits) and corresponding changes in fit quality (reduced $\chi^2$) and physical parameters (stellar mass, dust attenuation, and SFR) after refitting with DL-constrained AGN fractions $f_{\mathrm{AGN}}^{\mathrm{DL}}$ is shown. The exact values of $f_{\mathrm{AGN}}^{  \mathrm{DL}}$ are indicated in the top panel.}
    \label{fig: refit}
\end{figure}

The both-negligible region (Region A, green) is the most densely populated area, as evidenced by the transparency of the grey points. This indicates strong agreement between CIGALE and GRAHSP regarding negligible AGN contributions in these galaxies.
The mixed-detection region comprises galaxies 
with discrepant AGN contributions between CIGALE and GRAHSP. Within this region, GRAHSP AGN-significant galaxies (Region B, yellow) represent the majority (2,591 sources), while CIGALE AGN-significant galaxies (Region C, red) form a smaller subset (615 sources).
These categories account for $\sim12.1$\% and $\sim2.9$\% of the parent sample, respectively, representing an 80.8\% to 19.2\% split within the inconsistent population. The high level of agreement for galaxies with negligible AGN contribution, together with the larger discrepancies for candidate AGN-host systems, is consistent with previous comparative studies \citep{pacifici2023art,papadopoulos2025comparative}.

We further investigate the both-significant region by applying three criteria to identify galaxies with consistent AGN fraction estimates (in terms of absolute and fractional differences):
\begin{align}
\frac{|f_{\mathrm{AGN}}^{\mathrm{C}} - f_{\mathrm{AGN}}^{\mathrm{G}}|}
     {f_{\mathrm{AGN}}^{\mathrm{C}}} < 0.2, \\
\frac{|f_{\mathrm{AGN}}^{\mathrm{C}} - f_{\mathrm{AGN}}^{\mathrm{G}}|}
     {f_{\mathrm{AGN}}^{\mathrm{G}}} < 0.2, \\
\bigl| f_{\mathrm{AGN}}^{\mathrm{C}} - f_{\mathrm{AGN}}^{\mathrm{G}} \bigr| < 0.1.
\end{align}
Galaxies that satisfy all three criteria exhibit good agreement between $f_{\mathrm{AGN}}^{\mathrm{C}}$ and $f_{\mathrm{AGN}}^{\mathrm{G}}$ and are marked as Region E (blue). The remaining objects are labelled as Region~D (orange), indicating significant scatter. These two categories contain 233 and 348 galaxies, corresponding to $\sim1.1\%$ and $\sim1.6\%$ of the parent sample, and $\sim40.1\%$ and $\sim59.9\%$ of the both-significant population, respectively. Within the $\sim1.6\%$ scattered sub-sample, we find 286 ($\sim1.3\%$) galaxies with larger $f_{\mathrm{AGN}}^{\mathrm{G}}$ than $f_{\mathrm{AGN}}^{\mathrm{C}}$, and 62 ($\sim0.3\%$) galaxies for the reversed case, indicating a systematic tendency for GRAHSP to report higher AGN fractions.

\subsection{Comparison between DL and SEDs}
The comparison between CIGALE and GRAHSP reveals good agreement for galaxies with negligible AGN contributions, though discrepancies increase in AGN-dominated systems.
We observe a similar trend when comparing DL-based AGN fractions to the sample where both SED methods already agree. For galaxies with negligible AGN activity, consistency between DL and SED results remains high ($\sim86.4$\%), yielding an overall agreement of 71.1\% for the negligible-AGN sample (Fig.~\ref{fig:processing}). This consistency provides confidence in our methodology.

We then incorporate the $f_{\mathrm{AGN}}^{\mathrm{DL}}$ estimates, comparing them against the both-significant SED subsample (Regions D + E). To ensure a robust comparison within these AGN-significant sources, we apply a signal-to-noise cut on the SED-derived AGN fractions:

\begin{flalign}
        \label{eq:sigSED} &\ S/N (f_{\mathrm{AGN}}^{\mathrm{SED}}) > 1.  
\end{flalign}
This requirement ensures that the uncertainty does not exceed  the measured value itself. Applying this selection refines the both-significant subsample from 581 to 396 galaxies.

For this high-confidence AGN subsample ($N=396$ galaxies),  Fig.~\ref{fig: comparison} presents a detailed comparison of  
DL-derived $f_{\mathrm{AGN}}^{\mathrm{DL}}$ with CIGALE-derived $f_{\mathrm{AGN}}^{\mathrm{C}}$ (left column) 
and GRAHSP-derived $f_{\mathrm{AGN}}^{\mathrm{G}}$ (right column). 
Region D galaxies (scattered SED agreement) are marked with triangles; Region E galaxies (consistent SED agreement) are marked with circles.
To identify AGN significance in DL predictions, we adopt a threshold of $f_{\mathrm{AGN}}^{\mathrm{DL}} > 0.06$ from \citetalias{margalef2024agn}, corresponding to the $5\sigma$ threshold of the DL network. This differs from the $f_{\mathrm{AGN}}^{\mathrm{SED}} > 0.1$ criterion used for SED-based identifications. Under this threshold ($f_{\mathrm{AGN}}^{\mathrm{DL}} > 0.06$), only 30 galaxies ($\sim 7.6\%$) are identified as DL-significant among the 396 AGN-significant galaxies from the SED results, representing $\sim0.1\%$ of the parent sample (marked as coloured symbols). These 30 galaxies consist of 16 from Region D and 14 from Region E as defined in Fig.~\ref{fig: comparison_sed}. In Fig.~\ref{fig: AGN} and Fig.~\ref{fig: noAGN}, we show the galaxy imaging and SED-fitting results for four randomly selected AGN-significant and AGN-negligible galaxies, as identified by the DL model and both SED codes. The corresponding parameters are listed in Tab.~\ref{tab:galaxy_properties}.

These 30 DL-identified AGN generally scatter around the one-to-one relation  (especially in $f_{\mathrm{AGN}}^{\mathrm{C}}$ vs. $f_{\mathrm{AGN}}^{\mathrm{DL}}$). 
To quantify the level of agreement between SED and DL estimates and identify galaxies with consistent AGN fractions, we apply three consistency criteria analogous to those used for CIGALE–GRAHSP comparisons:
\begin{align}
\frac{|f_{\mathrm{AGN}}^{\mathrm{SED}} - f_{\mathrm{AGN}}^{\mathrm{  DL}}|}
     {f_{\mathrm{AGN}}^{\mathrm{DL}}} < 0.2, \\
\frac{|f_{\mathrm{AGN}}^{\mathrm{SED}} - f_{\mathrm{AGN}}^{\mathrm{  DL}}|}
     {f_{\mathrm{AGN}}^{\mathrm{SED}}} < 0.2,\\
\bigl| f_{\mathrm{AGN}}^{\mathrm{SED}} - f_{\mathrm{AGN}}^{\mathrm{  DL}} \bigr| < 0.1,
\end{align}
where the SED refers to both CIGALE and GRAHSP. These criteria define the shaded region in Fig.~\ref{fig: comparison} and Fig.~\ref{fig: comparison_ABC}.
As shown in the left panel of Fig~\ref{fig: comparison}, we identify ten galaxies with excellent value agreement between CIGALE and DL methods, while the remaining 20 galaxies exhibit larger systematic deviations. Among the well-matched cases, three belong to Region E.
In right panel, only four galaxies show good agreement between GRAHSP and DL, whereas the remaining 26 exhibit larger deviations; two of the four well-matched systems are located in Region E.

Beyond the high-confidence AGN detections in Regions D and E, we examine consistency between DL and SED estimates for the larger population of mixed-detection and AGN-negligible galaxies (Regions A, B, C; Fig.~\ref{fig: comparison_ABC}).
Triangular markers at $f_{\mathrm{AGN}}^{\mathrm{SED}}<0.1$ denote galaxies from Region A, for which a random $\sim$10\% subsample is displayed to reduce overcrowding. Most Region A galaxies ($\sim$86.5\%) remain AGN-negligible in the DL analysis, indicating good methodological agreement in identifying pure galaxy systems. However, a notable fraction ($\sim13.5$\%) exhibit possible AGN signatures in the DL imaging analysis despite SED classifications of negligible AGN. 
Among the mixed-detection galaxies (Regions B and C), we find partial overlap with DL AGN identifications: $\sim10.3\%$ of GRAHSP-significant galaxies are also identified as potential AGN hosts in the DL analysis, compared to $\sim9.4\%$ found for CIGALE-significant galaxies.

\section{Discussion}
\label{sec: discussion}
In Sect.~\ref{sec: results}, we quantitatively assessed the level of agreement between the DL- and SED-based $f_{\mathrm{AGN}}$ estimates. While the two approaches show broad consistency for AGN-negligible systems, discrepancies become substantial for AGN-significant systems. 
These findings suggest that directly incorporating imaging-based $f_{\mathrm{AGN}}$ constraints into SED fitting without proper reconciliation could introduce systematic biases.

In this section, we shift focus to understand when and why these discrepancies arise. We investigate the subsamples showing substantial discrepancies between the DL- and SED-based estimates, and assess which method provides  more reliable $f_{\mathrm{AGN}}$ measurements under different circumstances.
\subsection{AGN-significant galaxies: CIGALE refitting with $f_{\mathrm{AGN}}^{\mathrm{DL}}$}
\label{refit}

\begin{figure*}
    \centering
    \includegraphics[width=1.0\textwidth]{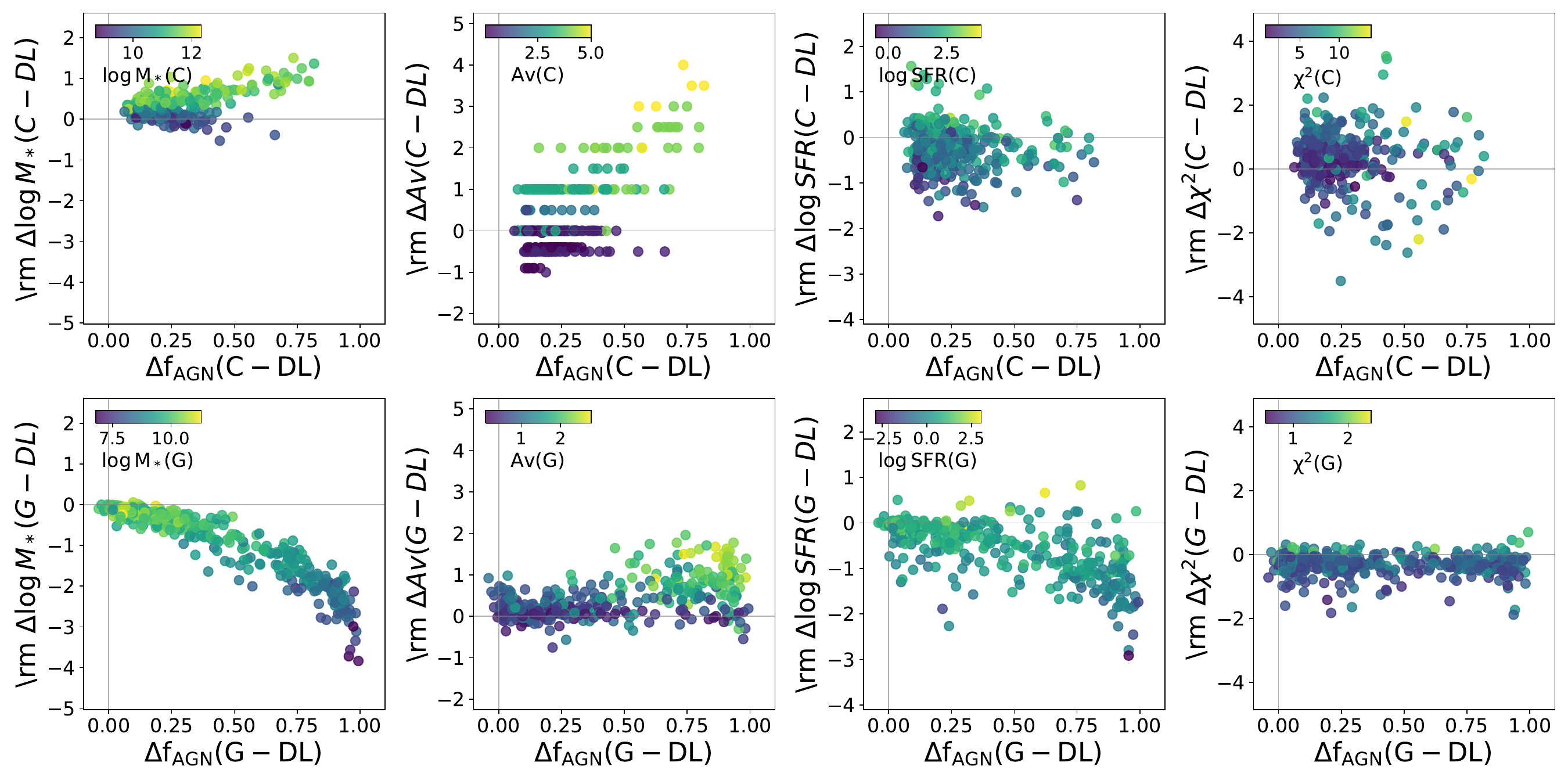}
    \caption{For galaxies with significant $f_{\mathrm{AGN}}^{\mathrm{SED}}$ ($>0.1$) but negligible $f_{\mathrm{AGN}}^{\mathrm{DL}}$ ($<0.06$), we show the difference between $f_{\mathrm{AGN}}^{\mathrm{SED}}$ and $f_{\mathrm{AGN}}^{\mathrm{DL}}$ as a function of the corresponding changes in physical parameters derived from SED fitting under the two assumptions. The upper row shows the results from CIGALE, and the lower row shows those from GRAHSP.}
    \label{fig: Param_difference_combined}
\end{figure*}

Among the 30 DL-significant galaxies above the $5\sigma$ threshold in Fig.~\ref{fig: comparison}, 20 exhibit moderate scatter between $f_{\mathrm{AGN}}^{\mathrm{DL}}$ and $f_{\mathrm{AGN}}^{\mathrm{C}}$.
To evaluate whether SED fitting can yield physically reasonable results when constrained by DL predictions, we refit these galaxies using CIGALE while fixing $f_{\mathrm{AGN}}^{\mathrm{C}} = f_{\mathrm{AGN}}^{\mathrm{DL}}$.
In the CIGALE configuration, we set the AGN fraction to the DL-predicted value and ensure the rest-frame wavelength range matches the JWST/NIRCam F150W band at each galaxy's redshift. We perform this analysis exclusively with CIGALE because GRAHSP does not support fixing AGN contribution fractions during SED fitting.

Figure~\ref{fig: refit} illustrates the differences in reduced $\chi^2$ and key physical parameters  between the original fits (open symbols) and the refitted results (black-edged symbols), colour-coded by 
$f_{\mathrm{AGN}}^{\mathrm{DL}}$. 
Despite a maximum variation of 0.4 in $f_{\mathrm{AGN}}^{\mathrm{DL}}$, 
AGN fraction constraints produce generally only modest changes in derived physical parameters and goodness-of-fit: maximum variations are 1.27 dex (stellar mass), 0.69 dex (dust attenuation), 0.81 dex (SFR), and 0.52 dex (reduced $\chi^2$).
While adopting $f_{\mathrm{AGN}}^{\mathrm{DL}}$ leads to a slight increase in $\chi^2_\nu$,  the resulting fits remain 
physically plausible and consistent with the observed photometry. The corresponding refitting SED-fitting results and parameters are shown in Fig.~\ref{fig: AGN} and listed in Table~\ref{tab:galaxy_properties}.

This finding indicates that 
AGN fraction variations have limited impact on SED-derived physical parameters in the observed NIR regime over $0.5 < z < 3$, consistent with the well-known degeneracy between AGN and host galaxy emission in the IR bands. This therefore highlights the value of incorporating imaging information into AGN fraction estimation and AGN--host decomposition.

\subsection{Galaxies with significant $f_{\mathrm{AGN}}^{\mathrm{SED}}$ but negligible $f_{\mathrm{AGN}}^{\mathrm{DL}}$}

As shown in Fig.~\ref{fig:processing}, the majority of SED-identified AGN-significant galaxies exhibit negligible $f_{\mathrm{AGN}}^{\mathrm{DL}}$ (366 galaxies, $\sim92.4\%$ after the S/N selection). While the SED decomposition for these galaxies suggests a dominant AGN emission contribution in the JWST/NIRCam F150W band, the corresponding imaging lacks a discernible central point source. 
To understand this systematic difference, we refit these 366 galaxies without an AGN component in both CIGALE and GRAHSP, and measure their galaxy sizes. In the following subsection, we examine $\Delta f_{\mathrm{AGN}}$ (difference between $f_{\mathrm{AGN}}^{\mathrm{SED}}$ and $f_{\mathrm{AGN}}^{\mathrm{DL}}$) as a function of physical parameters. We also make a comparison between the physical parameters derived from our SED fitting against those reported in the COSMOS2020 catalogue as an independent validation, and explore how different parameter combinations populate the stellar mass--size diagram.

\subsubsection{SED fitting without AGN component}
\label{sec:noAGN}
After refitting with both CIGALE and GRAHSP without an AGN component (guided by the $f_{\mathrm{AGN}}^{\mathrm{DL}}$ predictions), we analysed the resulting shifts in physical parameters as a function of $\Delta f_{\mathrm{AGN}}$.
The upper row of Fig.~\ref{fig: Param_difference_combined} displays the CIGALE refitting results. 
In the first column, we observe diverging trends in stellar mass. 
For low-mass galaxies, the stellar mass tends to increase with $\Delta f_{\mathrm{AGN}}$ (log$\rm M_*(C)$ < log$\rm M_*(DL)$, whereas a reversed trend is seen in high-mass galaxies. 
Dust attenuation, as a correlated parameter, exhibits a similarly contrasting behaviour, with $A_V$ increasing in low-$A_V$ galaxies and decreasing in high-$A_V$ galaxies.
Most refitted galaxies show an increased SFR when the AGN component is excluded, suggesting that a larger fraction of the total emission is being attributed to star formation. Interestingly, the majority of these galaxies show improved fit quality after removing the AGN component, even though \texttt{AGN fraction} = 0 was already an available option in the original configuration.
The lower row of Fig.~\ref{fig: Param_difference_combined} presents the GRAHSP refitting results. 
Unlike CIGALE, GRAHSP yields a monotonically increasing trend in physical parameters with 
$\Delta f_{\mathrm{AGN}}$, most notably in stellar mass and dust attenuation.
While more scattered,  SFR follows a similar upward trend. However, in terms of fit quality, nearly all GRAHSP galaxies show degraded residuals after AGN removal.
These different behaviours likely reflect differences in how the two codes parameterise  AGN contribution. CIGALE directly constrains the AGN fraction, whereas GRAHSP  regulates the overall AGN luminosity range and adopts different dust assumptions, leading to different degeneracies between parameters. These methodological differences highlight the need for caution when interpreting AGN fractions from SED fitting, particularly for systems with marginal AGN signatures, and underscore the value of independent imaging constraints.

\begin{figure}
    \centering
    \includegraphics[width=0.49\textwidth]{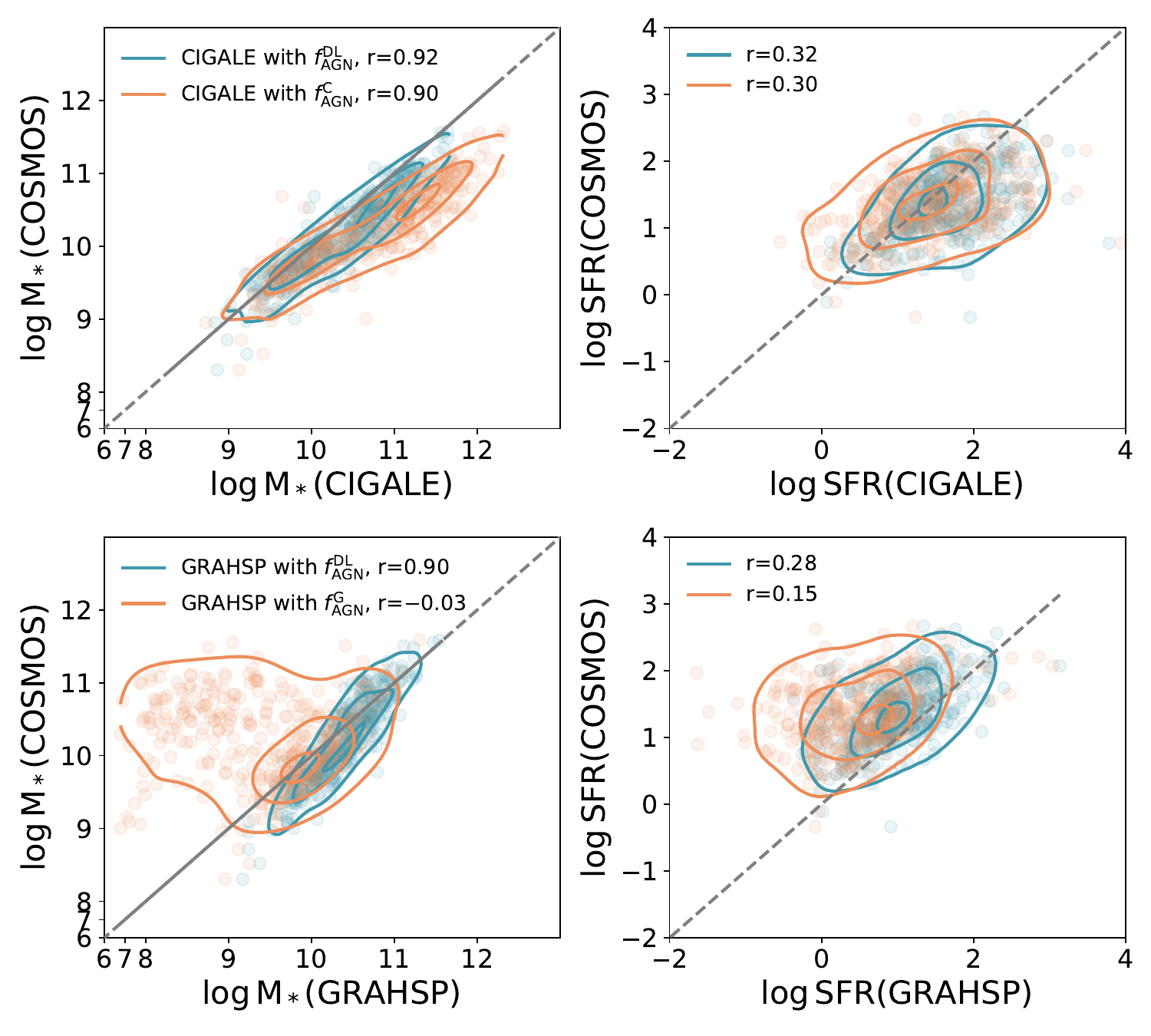}
    \caption{Galaxies with significant $f_{\mathrm{AGN}}^{\mathrm{SED}}$ ($>0.1$) but negligible $f_{\mathrm{AGN}}^{\mathrm{DL}}$ ($<0.06$), we show the comparison between stellar mass and star formation rate derived from SED fitting and those reported in the COSMOS2020 catalogue. The upper row shows results from CIGALE, while the lower row corresponds to GRAHSP. The blue contours with scatter points represent the fits without an AGN component, whereas orange contours and points denote fits including an AGN component. The dashed lines show the one-to-one relation, and the goodness of agreement is quantified by the Pearson correlation coefficient (r), as listed in the legend.}
    \label{fig:Param_difference_cosmos}
\end{figure}

Figure~\ref{fig:Param_difference_cosmos} places these parameter shifts in the context of the widely used COSMOS2020 catalogue. For both CIGALE and GRAHSP, fits without an AGN component generally lie closer to the COSMOS2020 values, while the inclusion of an AGN component introduces systematic offsets of varying magnitude. For stellar mass, CIGALE tends to increase the masses of high-mass galaxies, whereas GRAHSP yields a stronger decrease for low-mass systems. For SFR, both codes generally produce lower values when an AGN component is included. Overall, this comparison illustrates the scale of parameter shifts introduced by AGN modelling relative to a commonly adopted literature reference.

\subsubsection{Stellar mass--size diagram}

The stellar mass--size relation provides key insights into the assembly history of galaxies and their evolution over cosmic time. In this relationship, galaxy size is typically defined as the effective radius ($R_e$) of the stellar component. 
Consequently, properly accounting for potential AGN emission within the galaxy image is essential for obtaining reliable structural measurements.

To evaluate how different AGN fraction assumptions influence the stellar mass–size plane, we measure structural properties of our sample using \textsc{GALFIT} under two distinct scenarios: a pure galaxy model (i.e. negligible AGN) and a galaxy+AGN decomposition. 
In the AGN-negligible scenario, each galaxy is fitted with a single Sérsic profile. In contrast, in the AGN-dominated scenario, the galaxy image is decomposed into two components: a Sérsic component representing the stellar emission and a PSF component accounting for the AGN contribution. The flux of the PSF component is fixed according to the AGN fraction derived from SED fitting ($f_{\mathrm{AGN}}^{\mathrm{C}}$ or $f_{\mathrm{AGN}}^{\mathrm{G}}$) or DL ($f_{\mathrm{AGN}}^{\mathrm{DL}}$). 
As a result, for each galaxy we obtain three independent sets of  measurements: ($R_e^{\mathrm{DL}}$, $M_*^{\mathrm{DL}}$), ($R_e^{\mathrm{C}}$, $M_*^{\mathrm{C}}$), and ($R_e^{\mathrm{G}}$, $M_*^{\mathrm{G}}$).

Following \cite{weaver2023cosmos2020}, we separated quiescent galaxies (QGs) from star-forming galaxies (SFGs) using the NUV--r--J colour-colour diagram \citep{ilbert2013mass}:
\begin{equation}
\begin{aligned}
(NUV - r) > 3 (r - J), \\
(NUV - r) > 3.1,
\end{aligned}
\end{equation}
where the galaxy colour information is collected from the COSMOS2020 catalogue. Applying these criteria yields a sample of 338 SFGs and 28 QGs. Given the small number of QGs, we restrict the following analysis to SFGs only. We note, however, that this sample is re-selected from the parent sample without application of additional stellar mass-completeness criteria, which may bias the results toward higher masses. 

\begin{figure}
    \centering
    \includegraphics[width=0.495\textwidth]{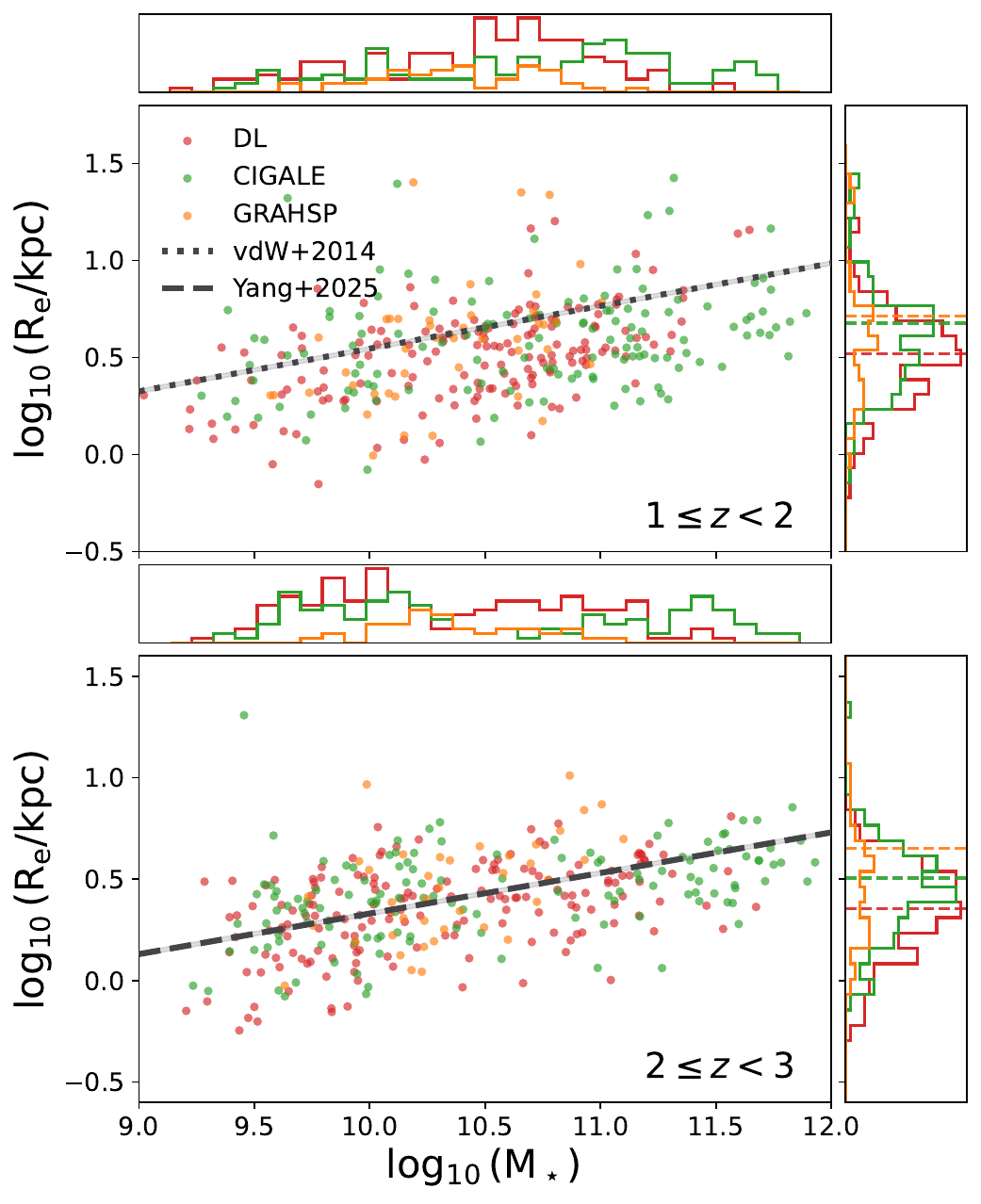}
    \caption{Stellar mass--size plane of SFGs in galaxies with significant $f_{\mathrm{AGN}}^{\mathrm{SED}}$ ($>0.1$) but negligible $f_{\mathrm{AGN}}^{\mathrm{DL}}$ ($<0.06$). The data points indicate the $M_*$ and $R_e$ derived from the DL, CIGALE, and GRAHSP methods. For comparison, the dotted line shows the reference relationship from HST \citep{van20143d} at $1 < z < 1.5$, whereas  the dashed line shows the reference relationship from COSMOS-Web \citep{shuntov2025cosmos} at $2 < z < 3$. 
    The histograms in the top and right panels present the one-dimensional distributions of stellar mass and effective radius, respectively.}
    \label{fig:msrelationship}
\end{figure}
We then plot these results on the stellar mass–size plane across two redshift bins ($1 < z < 2$ and $2 < z < 3$), as shown in Fig.~\ref{fig:msrelationship}. The red, green, and orange dots (and their corresponding histograms) represent measurements derived from the DL, CIGALE and GRAHSP methods, respectively. For comparison, the dotted and dashed lines denote reference scaling relations from HST \citep{van20143d} and COSMOS-Web \citep{shuntov2025cosmos} in similar redshift bins. 
As illustrated in Fig.~\ref{fig:msrelationship}, adopting different AGN contribution fractions can introduce systematic biases in the stellar mass-size plane. Both stellar mass and size estimates vary between methods. In particular, the effective radius $R_e$ tends to be smaller when lower $f_{\mathrm{AGN}}$ values are assumed.

\begin{figure*}
    \centering
    \includegraphics[width=1.0\textwidth]{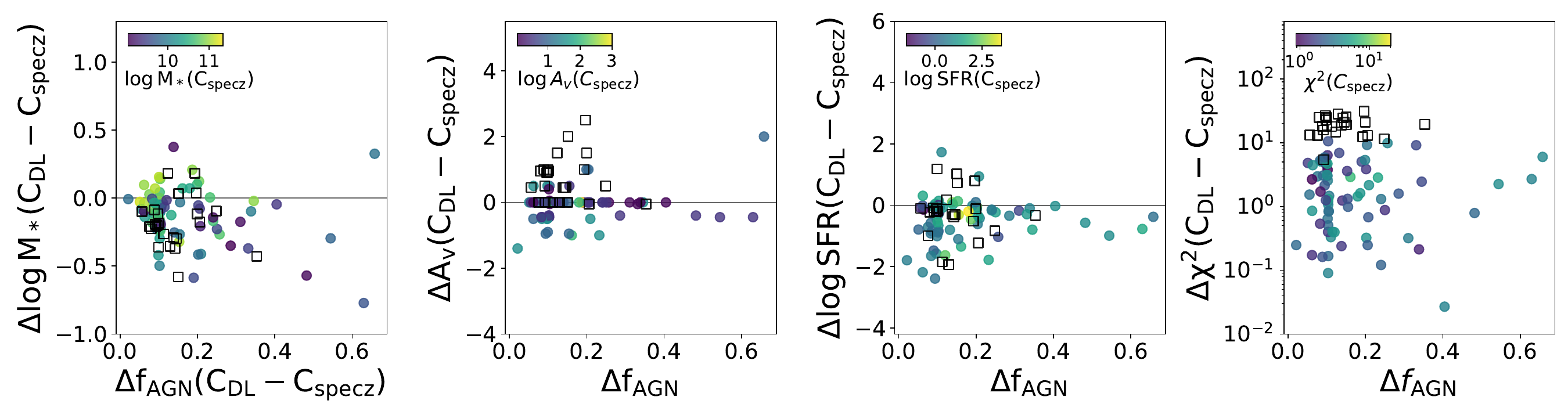}
    \caption{For the 88 galaxies with significant $f_{\mathrm{AGN}}^{\mathrm{DL}}$ ($>0.06$) but negligible $f_{\mathrm{AGN}}^{\mathrm{SED}}$ ($<0.1$), we perform two spec-z-based refits: a spec-z-only fit (\texttt{C\_specz}), and a spec-z fit with an additional constraint $f_{\mathrm{AGN}}^{\mathrm{C\_DL}} = f_{\mathrm{AGN}}^{\mathrm{DL}}$ (\texttt{C\_DL}). The colour bar encodes the \texttt{C\_specz} parameter values. We show the difference in $f_{\mathrm{AGN}}$ as a function of the corresponding changes in physical parameters between these two fits. Empty squares indicate failed fits.}
    \label{fig:Param_difference_newZ}
\end{figure*}

\subsection{Galaxies with significant $f_{\mathrm{AGN}}^{\mathrm{DL}}$ but negligible $f_{\mathrm{AGN}}^{\mathrm{SED}}$}

While DL and SED methods show strong agreement for AGN-negligible galaxies, there remains a notable population of 2,397 sources ($\sim11.2$\%) classified as AGN-significant by the DL model. These systems occupy the top-left region of both panels in Fig.~\ref{fig: comparison_ABC}. For these galaxies, SED fitting suggests that the emission is almost entirely host-dominated. However, the JWST/NIRCam F150W imaging reveals a central point source contributing at least ~10\% of the total galaxy light. This raises important questions: are these genuine AGN missed by SED fitting, or are they compact structures (e.g., bulges, star clusters, or morphological clumps) misidentified as AGN by the DL network?

To investigate whether photometric redshift (photo-z) uncertainties contribute to these discrepancies,
we cross-matched our sample with the COSMOS Spectroscopic Redshift Compilation catalogue \citep{khostovan2025cosmos}. We identified 101 galaxies with redshift differences $>0.1$ whose spectroscopic redshifts lie within the range $0.5<z<3$. 
For these sources, we reran the analysis using corrected spectroscopic redshifts to determine whether improved redshifts resolve the AGN discrepancies.
Specifically, we performed two spec-$z$-based refits: 
\begin{itemize}
    \item \texttt{C\_specz}: Refitting only with the new spec-$z$.
    \item \texttt{C\_DL}:  Refitting with the new spec-$z$ and AGN fraction fixed to DL estimate $f_{\mathrm{AGN}}^{\mathrm{C\_DL}} = f_{\mathrm{AGN}}^{\mathrm{DL}}$.
\end{itemize}
As described before, we only utilise CIGALE for this refitting process.

After refitting in the first scenario (\texttt{C\_specz}), 88 galaxies retained negligible $f_{\mathrm{AGN}}$ values with improved redshifts, demonstrating that photo-$z$ errors explain only a few some  discrepant cases.
Results are shown in Fig.~\ref{fig:Param_difference_newZ}. The x- and y-axes represent the difference in AGN contribution and the resulting change in physical parameters between \texttt{C\_DL} and \texttt{C\_specz}, respectively. Additionally, the parameter values from the \texttt{C\_specz} run are represented by the colour-coding. 
Open squares highlight galaxies where the reduced $\chi^{2}$ from \texttt{C\_DL} exceeds the $5\sigma$ threshold defined in  Table~\ref{tab: chi2}. 
These are classified as failed fits, as the refitting  (constrained by $f_{\mathrm{AGN}}^{\mathrm{C\_DL}} = f_{\mathrm{AGN}}^{\mathrm{DL}}$) fails to converge to a physically reasonable solution.
Around 30\% of the galaxies are categorised as failed fits. This indicates that the AGN signal detected by the DL model may instead originate from compact galaxy bulges, localised star-forming regions, or clumpy morphological structures. For the successfully refitted galaxies, we observe only a minor degradation in the reduced $\chi^{2}$ from \texttt{C\_DL}. Other physical parameters, including stellar mass, SFR, and dust attenuation, exhibit moderate variations, 
with no clear systematic correlation between the magnitude of these shifts and the overall success of the refit.

\section{Conclusions}
\label{sec: conclusions}

In this study, we developed a unified framework to quantify AGN contributions by combining three independent approaches: SED fitting with CIGALE and GRAHSP, and imaging-based  decomposition from \citet{margalef2024agn}. Utilising multi-wavelength photometry from COSMOS2020, supplemented by JWST photometry from COSMOS2025, Chandra X-ray  detections, and deblended FIR measurements, we applied this framework to a mass-complete sample refined by SED-fitting quality cuts. For these 21,428 galaxies, we systematically compared how each method identifies significant AGN activity and evaluated the robustness and inherent limitations of SED-derived AGN fractions. 
The main findings of this work are summarised as follows:

\begin{enumerate}
  \item Strong agreement for AGN-negligible galaxies. The three AGN identification methods demonstrate remarkable consistency for low-AGN systems. Specifically, 82.4\% of the entire sample is identified as AGN-negligible by both CIGALE and GRAHSP, with 86.4\% of these cases also showing a minimal AGN contribution according to  the DL  imaging analysis. This high concordance suggests that SED fitting and imaging methods reliably identify pure galaxy systems. 
  
  \item 
  Discrepancies increase systematically for AGN-dominant scenarios. When comparing $f_{\mathrm{AGN}}$ from the two SED codes, we find that 15.0\% of the sources are inconsistent between CIGALE and GRAHSP. Strikingly, in four-fifths of these inconsistent cases, GRAHSP identifies a significant AGN component where CIGALE does not, indicating a systematic tendency for GRAHSP to favor higher AGN fractions. Only 2.7\% of the galaxies are classified as AGN-significant by both SED codes. Within this subset, only 40\% show tight quantitative agreement, while the remaining 60\% exhibit substantial scatter, indicating uncertain AGN characterisation even when both codes agree on AGN presence.

  \item 
  Agreement remains limited when incorporating the DL predictions. Among galaxies identified as AGN-significant by both SED fitting, only 7.6\% are consistently classified as hosting a significant AGN by the DL method. Furthermore, within this small subset, only ten galaxies exhibits robust quantitative agreement across all three methods.

    \item 
Re-running CIGALE by fixing $f_{\mathrm{AGN}}$ to the DL-estimated value for 20 AGN-dominant galaxies yielded statistically acceptable fits, with a maximum reduced $\chi^2$
 shift of 0.52 dex. Fixing the $f_{\mathrm{AGN}}$
 value resulted in maximum parameter variations of 1.27 dex in stellar mass, 0.69 dex in attenuation, and 0.81 dex in SFR.

    \item For galaxies with significant $f_{\mathrm{AGN}}^{\mathrm{SED}}$ but negligible $f_{\mathrm{AGN}}^{\mathrm{DL}}$, we obtained physically acceptable results by refitting the SEDs without an AGN component. 
    However, the decision to adopt either a dominant or negligible $f_{\mathrm{AGN}}$ can introduce systematic biases in the resulting stellar mass–size plane.

    \item For galaxies with negligible $f_{\mathrm{AGN}}^{\mathrm{SED}}$ but significant $f_{\mathrm{AGN}}^{\mathrm{DL}}$, we find that $\sim$30\% fail to yield a physically plausible  solution when refitted with $f_{\mathrm{AGN}}^{\mathrm{C\_DL}}$, as evidenced by their elevated reduced $\chi^2$. 
    Furthermore, we observe no clear correlation between the success of these refits and the magnitude of the resulting shifts in physical parameters.

   \end{enumerate}

Discrepancies between SED-based and imaging-based AGN fraction estimates highlight the need for a unified framework to leverage the complementary strengths of both approaches. While SED fitting exploits broad multi-wavelength photometric coverage and physically motivated models, its estimates are often limited by parameter degeneracies arising from the imbalance between the number of free parameters and the constraining power of the data. This can lead to inferred AGN contributions that are inconsistent with the observed source morphology. Incorporating information from spatially resolved  imaging data offers a promising way to mitigate these degeneracies, particularly those associated with dust attenuation and AGN--stellar emission mixing.

Practically, based on our results, we recommend a joint diagnostic framework for AGN characterisation that compares at least one SED-based estimate with imaging-based morphological decomposition. For galaxies where SED fitting and imaging decomposition are consistent, the majority are found to be AGN-negligible, with only a small fraction showing strong AGN contributions. 
These concordant cases represent robust AGN identifications requiring no further scrutiny.
For galaxies where SED and imaging methods disagree, we propose the following classification and treatment protocols:
\begin{itemize}
    \item Galaxies with significant $f_{\mathrm{AGN}}^{\mathrm{SED}}$ but negligible $f_{\mathrm{AGN}}^{\mathrm{DL}}$. These systems likely represent "SED AGN" artifacts arising from decomposition ambiguities rather than genuine AGN. Therefore, in these cases, we recommend prioritising the imaging-based classification (i.e. negligible AGN). To validate this, SED fitting can be rerun without an AGN component. If the resulting fit remains acceptable, the non-AGN solution and the corresponding physical parameters should be adopted. This approach mitigates SED parameter degeneracies that artificially inflate inferred AGN fractions.
    \item For galaxies with negligible $f_{\mathrm{AGN}}^{\mathrm{SED}}$ but significant $f_{\mathrm{AGN}}^{\mathrm{DL}}$. These systems may harbor genuine AGN missed by SED decomposition or represent compact non-AGN structures misidentified as AGN by imaging decomposition. We recommend rerunning SED fitting with the AGN fraction explicitly constrained to the imaging-based estimate, i.e. fixing $f_{\mathrm{AGN}} = f_{\mathrm{AGN}}^{\mathrm{DL}}$. If a physically reasonable solution with acceptable fit quality is obtained, the imaging-based AGN contribution should be adopted. If  SED fitting fails to converge or yields unphysical results, these galaxies should be flagged as special cases warranting further investigation with higher-resolution imaging or complementary diagnostics (e.g., X-ray, radio or  spectroscopy).
\end{itemize}
The rapid growth of high-resolution imaging from facilities such as JWST/NIRCam and \textit{Euclid} (VIS and NISP YJH bands) enables more robust joint analyses that unify SED decomposition with imaging-based constraints. Adopting such integrated approaches will be essential for deriving reliable AGN contribution estimates across cosmic time.

\begin{acknowledgements}
      This publication is part of the project `Clash of the Titans: deciphering the enigmatic role of cosmic collisions' (with project number VI.Vidi.193.113 of the research programme Vidi, which is (partly) financed by the Dutch Research Council (NWO). We thank Johannes Buchner and  Mara Salvato for help in organising the paper and utilising the GRAHSP code. We thank Mederic Boquien for help in adjusting the CIGALE code. The SED-fitting process was carried out on Hábrók. We thank the Centre for Information Technology of the University of Groningen for their support and for providing access to the Hábrók high performance computing cluster. We thank SURF (www.surf.nl) for the support in using the National Supercomputer Snellius.
      We acknowledges support from China Scholarship Council.
      K.M. acknowledges support from the grant UMO-2024/53/B/ST9/00230 funded by the National Science Centre, Poland.
       This research was supported by the Polish National Agency for Academic Exchange under the CLEVER, Strategic Partnership programme BPI/PST/2024/1/00019
       \end{acknowledgements}

\bibliographystyle{aa} 
\bibliography{ref}

\begin{appendix} 
\onecolumn 

\section{Photometric catalogue}
\label{app:catalogue}
In Tab.~\ref{tab: catalogue}, we report the photometric catalogue that used for the SED fitting. We note that the X-ray data are not included in the GRAHSP fits, as the current GRAHSP library does not include a corresponding X-ray module.
{
\small                
\renewcommand{\arraystretch}{1} 
\setlength{\tabcolsep}{1.5pt}  

\setlength{\LTcapwidth}{\textwidth}

\setlength{\LTleft}{\fill}
\setlength{\LTright}{\fill}

\newlength{\mytablewidth}
\setlength{\mytablewidth}{9.8cm}

\begin{longtable}{@{} 
>{\raggedright\arraybackslash}p{2.8cm} 
>{\centering\arraybackslash}p{1.5cm} 
>{\centering\arraybackslash}p{2cm} 
>{\raggedright\arraybackslash}p{3.2cm} 
>{\centering\arraybackslash}p{0.8cm} 
@{}}
\caption{X-ray--optical--IR data used in this study.}
\label{tab: catalogue}\\

\toprule 
\midrule
Facility & Band & Central & Depth & \\
\midrule
\endfirsthead
\endhead

    Chandra & 0.5–7 keV & - & $1.2 \times 10^{-15}\,\mathrm{erg}\,\mathrm{s}^{-1}\,\mathrm{cm}^{-2}$& \multirow{3}{*}{$\left\}\rule{0pt}{2em}\right.{a}$}  \\

    Chandra & 0.5–2 keV & - & $2.8 \times 10^{-16}\,\mathrm{erg}\,\mathrm{s}^{-1}\,\mathrm{cm}^{-2}$ & \\

    Chandra & 2–10 keV & - & $1.9 \times 10^{-15}\,\mathrm{erg}\,\mathrm{s}^{-1}\,\mathrm{cm}^{-2}$ & \\
    GALEX & FUV & 1526 \AA & 26.0 & \multirow{34}{*}{$\left\}\rule{0pt}{20.5em}\right.{b}$} \\
    GALEX & NUV & 2307 \AA & 26.0 & \\

    MegaCam/CFHT & u & 3709 \AA & 27.8 & \\
    MegaCam/CFHT & u$^*$ & 3858 \AA & 27.7 & \\
    ACS/HST & F814W & 8333 \AA & 27.8 & \\
    HSC/Subaru & g & 4847 \AA & 28.1 & \\
    HSC/Subaru & r & 6219 \AA & 27.8 & \\
    HSC/Subaru & i & 7699 \AA & 27.6 & \\
    HSC/Subaru & z & 8894 \AA & 27.2 & \\
    HSC/Subaru & y & 9761 \AA & 26.5 & \\
    Suprime-Cam/Subaru & IB427 & 4266 \AA & 26.1 & \\

    Suprime-Cam/Subaru & IB464 & 4635 \AA & 25.6 & \\
    Suprime-Cam/Subaru & IA484 & 4851 \AA & 26.5 & \\
    Suprime-Cam/Subaru & IB505 & 5064 \AA & 26.1 & \\
    Suprime-Cam/Subaru & IA527 & 5261 \AA & 26.4 & \\
    Suprime-Cam/Subaru & IB574 & 5766 \AA & 25.8 & \\
    Suprime-Cam/Subaru & IA624 & 6232 \AA & 26.4 & \\
    Suprime-Cam/Subaru & IA679 & 6780 \AA & 25.6 & \\
    Suprime-Cam/Subaru & IB709 & 7073 \AA & 25.9 & \\
    Suprime-Cam/Subaru & IA738 & 7361 \AA & 26.1 & \\
    Suprime-Cam/Subaru & IA767 & 7694 \AA & 25.6 & \\
    Suprime-Cam/Subaru & IB827 & 8243 \AA & 25.6 & \\
    Suprime-Cam/Subaru & NB711 & 7121 \AA & 25.5 & \\
    Suprime-Cam/Subaru & NB816 & 8150 \AA & 25.6 & \\

    VIRCAM/VISTA & Y & 10216 \AA & 26.6/25.3$^e$ & \\
    VIRCAM/VISTA & J & 12525 \AA & 26.4/25.2 & \\
    VIRCAM/VISTA & H & 16466 \AA & 26.1/24.9 & \\
    VIRCAM/VISTA & Ks & 21557 \AA & 25.7/25.3 & \\
    VIRCAM/VISTA & NB118 & 11909 \AA & 24.8 & \\

    IRAC/Spitzer & ch1 & 35686 \AA & 26.4 & \\
    IRAC/Spitzer & ch2 & 45067 \AA & 26.3 & \\
    IRAC/Spitzer & ch3 & 57788 \AA & 23.2 & \\
    IRAC/Spitzer & ch4 & 79958 \AA & 23.1 & \\

    \midrule

    NIRCam/JWST & F115W & 11622 \AA & 27.2 & \multirow{5}{*}{$\left\}\rule{0pt}{2.9em}\right.{c}$} \\
    NIRCam/JWST & F150W & 15106 \AA & 27.4 & \\
    NIRCam/JWST & F277W & 28001 \AA & 28.1 & \\
    NIRCam/JWST & F444W & 44366 \AA & 28.0 & \\
    MIRI/JWST & F770W & 77108 \AA & 25.2 & \\

    \midrule

    MIPS/Spitzer & 24$\mathrm{\mu m}$ & 23.59 $\mu$m & 109.4 $\rm \mu Jy$& \multirow{6}{*}{$\left\}\rule{0pt}{3.2em}\right.{d}$} \\
    PACS/Herschel & green & 100.65 $\mu$m & 7.0 $\rm mJy$& \\
    PACS/Herschel & red & 160.98 $\mu$m & 18.0 $\rm mJy$& \\
    SPIRE/Herschel & PSW & 248.29 $\mu$m & 35.0 $\rm mJy$& \\
    SPIRE/Herschel & PMW & 348.44 $\mu$m & 34.3 $\rm mJy$& \\
    SPIRE/Herschel & PLW & 500.38 $\mu$m & 32.4 $\rm mJy$& \\

\bottomrule

\end{longtable}

\vspace{-0.5em}

\noindent\textbf{Notes.} The columns list the facility, band, central wavelength, and observational depth. All listed magnitudes are in the AB system. 

\noindent\textsuperscript{a} X-ray data are drawn from \citep{la2026major}.

\noindent\textsuperscript{b} Data are taken from COSMOS2020 \citep{weaver2022cosmos2020}. The depths are mostly measured within a $2''$ aperture at the $3\sigma$ level.

\noindent\textsuperscript{c} Data are taken from COSMOS2025 \citep{shuntov2025cosmos2025}. The depths are calculated within $0.15''$ NIRCam/JWST and $0.5''$ MIRI/JWST apertures at $5\sigma$.

\noindent\textsuperscript{d} Data are drawn from the ancillary deblending catalogue \citep{wang2024probabilistic}. The $5\sigma$ depths are derived from the $1\sigma$ depths.

\noindent\textsuperscript{e} For the Y, J, H, and $\mathrm{K}_\mathrm{S}$ bands from VIRCAM/VISTA, the first depth refers to the ultra-deep field and the second to the deep field. We prioritise the ultra-deep data and use the deep-field data only in regions not covered by the ultra-deep observations.
}

\section{AGN library in SED fitting}
\label{app:library_diff}

\begin{figure}[h]
    \centering
    \includegraphics[width=0.975\linewidth]{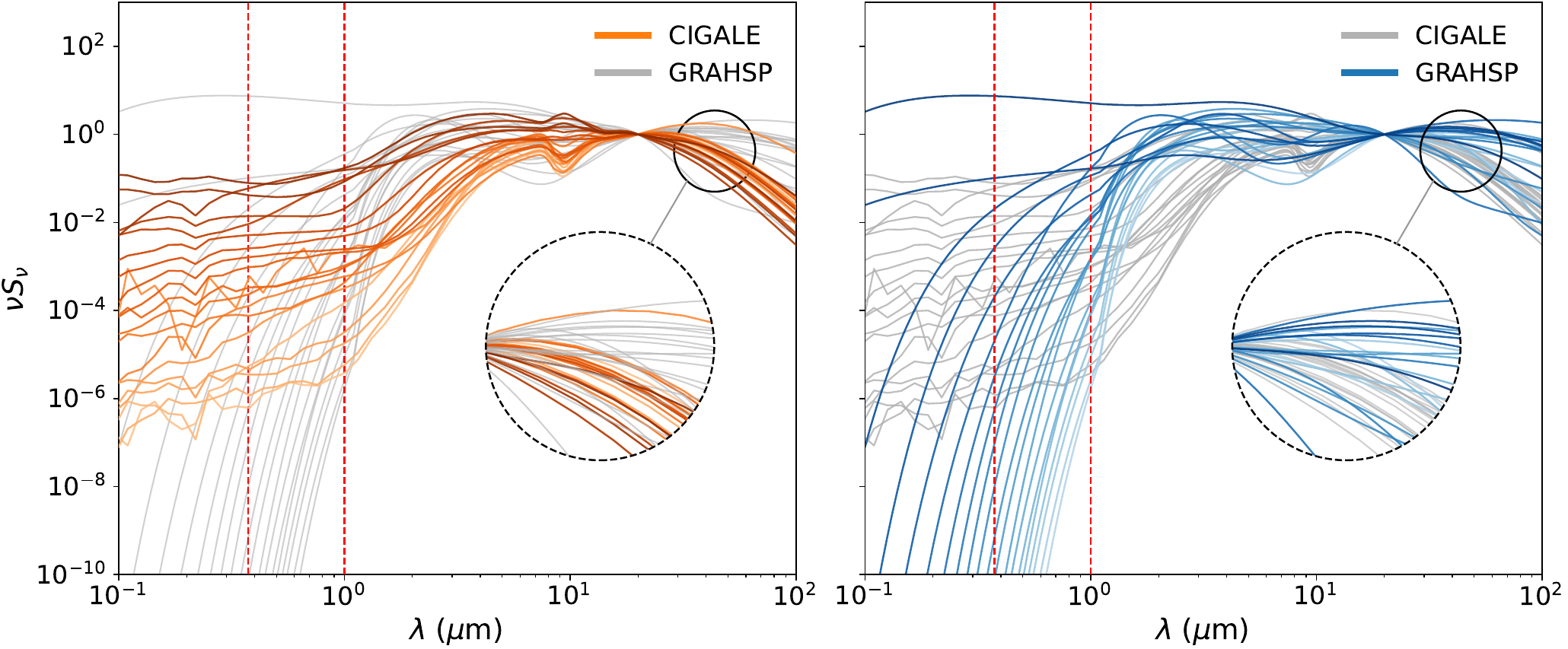}
    \caption{
    Comparison of the AGN template libraries used in CIGALE and GRAHSP.
    The left panel shows ten CIGALE AGN templates (orange) overlaid on the GRAHSP library (grey), while the right panel shows ten GRAHSP AGN templates (blue) overlaid on the CIGALE library (grey).
    All templates are normalised at $20\, \mu\mathrm{m}$, with $\nu S_\nu=1$.
    The red dashed lines at $1$ and $0.375\,\mu\mathrm{m}$ indicate the rest-frame wavelength range corresponding to the JWST/NIRCam F150W band over $z \sim 0.5$--3. 
   The inset provides an expanded view of the $20$--$60\,\mu\mathrm{m}$ range to highlight the MIR differences between the two models.
    }
    
    \label{fig: agn library} 
\end{figure}

Figure~\ref{fig: agn library} presents the AGN templates adopted in GRAHSP and CIGALE. 
For each code, we uniformly select ten representative templates spanning the full envelope of the AGN template library to illustrate the range of AGN model shapes allowed by each model. 
Although both libraries cover a broad parameter space, their template shapes differ systematically. 
Notably, GRAHSP templates extend to lower flux levels at shorter wavelengths compared to CIGALE. It may due to the more flexible treatment of hot-dust emission in the GRAHSP AGN model.

\section{SED module parameters}
Tab.~\ref{tab:cigale_parameters} summarises the modules and parameters adopted in CIGALE-fitting, whereas Tab.~\ref{tab:grahsp_parameters} presents the corresponding setup used for GRAHSP-fitting.

{\small
\setlength{\LTcapwidth}{\textwidth}

\begin{longtable}{lll}
\caption{Initial settings in the CIGALE runs.}
\label{tab:cigale_parameters}\\

\toprule
\midrule
\endfirsthead

\toprule
\midrule
\endhead

\bottomrule
\endlastfoot

  \multicolumn{3}{c}{\textbf{\large Star-Formation History}} \\
\midrule
  delayed $\tau$+starburst & e-folding time of the main population & 500, 1000-5000 (step 1000) Myrs \\
  & age the main population & 1000, 1500, $2000-8000$ (step 1000) Myrs \\
  & e-folding time of the late starburst population & 9000, 13000 Myrs \\
  & age of the late starburst population & 1, 50, 150 Myrs \\
  & mass fraction of the late starburst population & 0.0, 0.05, 0.1, 0.2\\
  \hline\\[-7pt]
  \multicolumn{3}{c}{\textbf{\large Single Stellar Population}}\\
\midrule
  \citet{bruzualStellarPopulationSynthesis2003} & IMF & Chabrier (1) \\
  & metallicity & solar (0.02) \\
\midrule
  \multicolumn{3}{c}{\textbf{\large Dust attenuation}}\\
\midrule
  modified & V-band attenuation in the interstellar medium & 0.05, 0.1, 0.5, 1, 1.5, 2, 3, 4, 5\\
  \citet{charlotSimpleModelAbsorption2000} & Av$_{ISM} $/ Av$_{BC}$+Av$_{ISM}$ & 0.25, 0.5, 0.75 \\
  & Power law slope of the attenuation in the ISM & $-0.7$ \\
  & Power law slope of the attenuation in the birth clouds & $-0.48$, $-0.7$ \\
\midrule
  \multicolumn{3}{c}{\textbf{\large Dust emission}}\\
\midrule
  \citet{draineAndromedasDust2014} & Mass fraction of PAH & 0.47, 1.12, 2.5 \\
  & minimum radiation field & 5, 10, 25\\
  & power-law slope $\alpha$ in dM/dU $\propto U^{-\alpha}$ & 2.0 \\
\midrule
  \multicolumn{3}{c}{\textbf{\large X-ray}}\\
\midrule
  X-CIGALE X-ray & AGN photon index $\Gamma$ & 1.8\\
  \citep{yang2020x, yang2022fitting} & power slope $\alpha_{ox}$ & -1.8, -1.6, -1.4 \\
    & Max deviation from the $\alpha_{ox}$--$L_{\nu, 2500\,\text{\AA}}$ relation & 0.4 \\
    & AGN X-ray angle coefficients & (0.5, 0) \\
\midrule
  \multicolumn{3}{c}{\textbf{\large AGN template}}\\
\midrule
  SKIRTOR & Average edge-on optical depth at 9.7 $\mu$m & 7 \\
  \citep{stalevski2016dust} & torus density radial parameter $p$ & 1 \\
  & torus density angular parameter $q$ & 1\\
  & Angle between the equatorial plane and edge of the torus & 40$\degree$ \\
  & viewing angle & 30$\degree$ (type 1), 70$\degree$ (type 2) \\
  & AGN fraction & 0.01, 0.05, 0.07, 0.1, 0.125, 0.15, 0.2, 0.25, 0.3, \\ & & 0.45, 0.55, 0.6, 0.7, 0.8 \\
  & extinction law of polar dust & SMC (0) \\
  & E(B-V) of polar dust & 0.2\\
  & temperature of polar dust & 100\\
  & emissivity of polar dust & 1.6\\
\midrule
  Fritz & Ratio of the maximum to minimum radii of the dust torus & 30.0 \\
  \citep{fritz2006revisiting} & Optical depth at 9.7 microns $p$ & 3.0, 6.0, 10.0 \\
  & radial dust distribution in the torus & -0.5\\
  & angular dust distribution in the torus & 4.0 \\
  & angular opening angle of the torus & 60$\degree$, 100$\degree$, 140$\degree$\\
  & angle between equatorial axis and the line of sight & 0.001$\degree$, 89.900$\degree$ \\
  & AGN fraction,  & 0, 0.1, 0.2, 0.3, 0.45, 0.60, 0.75, 0.9, 0.99 \\
  & extinction law of polar dust & SMC (0) \\
  & E(B-V) of polar dust & 0, 0.2, 0.4\\
  & temperature of polar dust & 100\\
  & emissivity of polar dust & 1.6\\

  \end{longtable}

\vspace{-0.5em}
\noindent\textbf{Notes.} 
The first column indicates the model utilised, the second column briefly describes the parameters customised for this analysis, and the third column lists the corresponding parameter set-ups.
}

{\small
\setlength{\LTcapwidth}{\textwidth}
\begin{longtable}{lll}
\caption{\noindent Initial settings in the GRAHSP runs.}
\label{tab:grahsp_parameters}\\

\toprule
\midrule
\multicolumn{3}{c}{\textbf{\large Galaxy components}} \\
\midrule
stellar mass &  & log-uniform between $10^5$ and $10^{15}$ $M_\odot$ \\
delayed $\tau$+starburst &  &  \\
\citet{bruzualStellarPopulationSynthesis2003} & the same as CIGALE runs &  \\
\citet{draineAndromedasDust2014} &  &  \\

\midrule
\multicolumn{3}{c}{\textbf{\large AGN components}} \\
\midrule
\texttt{$L_{\rm AGN}$}: AGN luminosity &  & log-uniform between $10^{38}$ and $10^{50}$ erg s$^{-1}$ \\

\midrule
blue bump continuum 
& \texttt{plslope}: power slope  
& uniform between $-2.7$ and $-1$ in steps of 0.01 \\

& \texttt{uvslope}: power slope of UV side  
& 0 \\

& \texttt{plbendloc}: break wavelength between two slopes  
& 50, 80, 90, 100, 120, 150 nm \\

& \texttt{plbendwidth}: width of the bend   
& log-uniform between 0.1 and 10 with 10 steps \\

\midrule
emission lines 
& \texttt{AFeII}: FeII / $H_\beta$ luminosity ratio  
& log-uniform between $-0.2$ and 1.5 with 10 steps \\

& \texttt{Alines}: scaling factor of the emission-line template 
& 0.3, 0.5, 0.7, 1, 1.5, 2, 4, 10, 20 \\

& \texttt{linewidth}: full width at half maximum  
& 10000 km s$^{-1}$ \\

\midrule
infrared torus 
& \texttt{Si}: absorption or emission from Si 
& uniform between $-4$ and $+4$ in steps of 0.2 \\  

& \texttt{fcov}: torus / UV--optical continuum amplitude ratio 
& uniform between 0.05 and 0.95 in steps of 0.05 \\

& \texttt{COOLlam}: peak wavelength of the cool dust emission  
& uniform between 10 and 30 in steps of 0.01 \\

& \texttt{COOLwidth}: width of the cool dust emission 
& uniform between 0.2 and 0.70 in steps of 0.05 \\

& \texttt{HOTlam}: peak wavelength of the hot dust emission  
& uniform between 1 and 5.5 in steps of 0.01 \\

& \texttt{HOTwidth}: width of the hot dust emission 
& uniform between 0.2 and 0.70 in steps of 0.05 \\

& \texttt{HOTfcov}: hot-to-cold dust peak normalisation ratio in $\lambda L_\lambda$ 
& 0.04, 0.1, 0.2, 0.4, 0.6, 0.8, 1.0, 1.2, 1.4, 1.6, 2.0, 2.5, \\
 & & 3, 5, 10 \\

\midrule
\multicolumn{3}{c}{\textbf{\large Attenuation}} \\
\midrule
bi-component attenuation 
& E(B$-$V) of host galaxy 
& log-uniform between 0.01 and 10 with 80 steps \\

& E(B$-$V) of AGN  
& log-uniform between 0.01 and 10 with 80 steps \\

\bottomrule
  \end{longtable}

\vspace{-0.5em}
\noindent\textbf{Notes.} 
The first column indicates the model utilised, the second column briefly describes the parameters customised for this analysis, and the third column lists the corresponding parameter set-ups.
}

\section{SED fitting results and galaxy imaging for the AGN-significant and -negligible samples}
Table~\ref{tab:galaxy_properties} presents the SED-fitting parameters for 30 AGN-significant galaxies identified by CIGALE, GRAHSP, and DL. For each physical parameter, we report results from both SED methods. Values in brackets correspond to the CIGALE refits with $f_{\mathrm{AGN}}$ fixed to the DL estimates.

\begin{figure*}[h]
\centering
\includegraphics[width=0.975\textwidth]{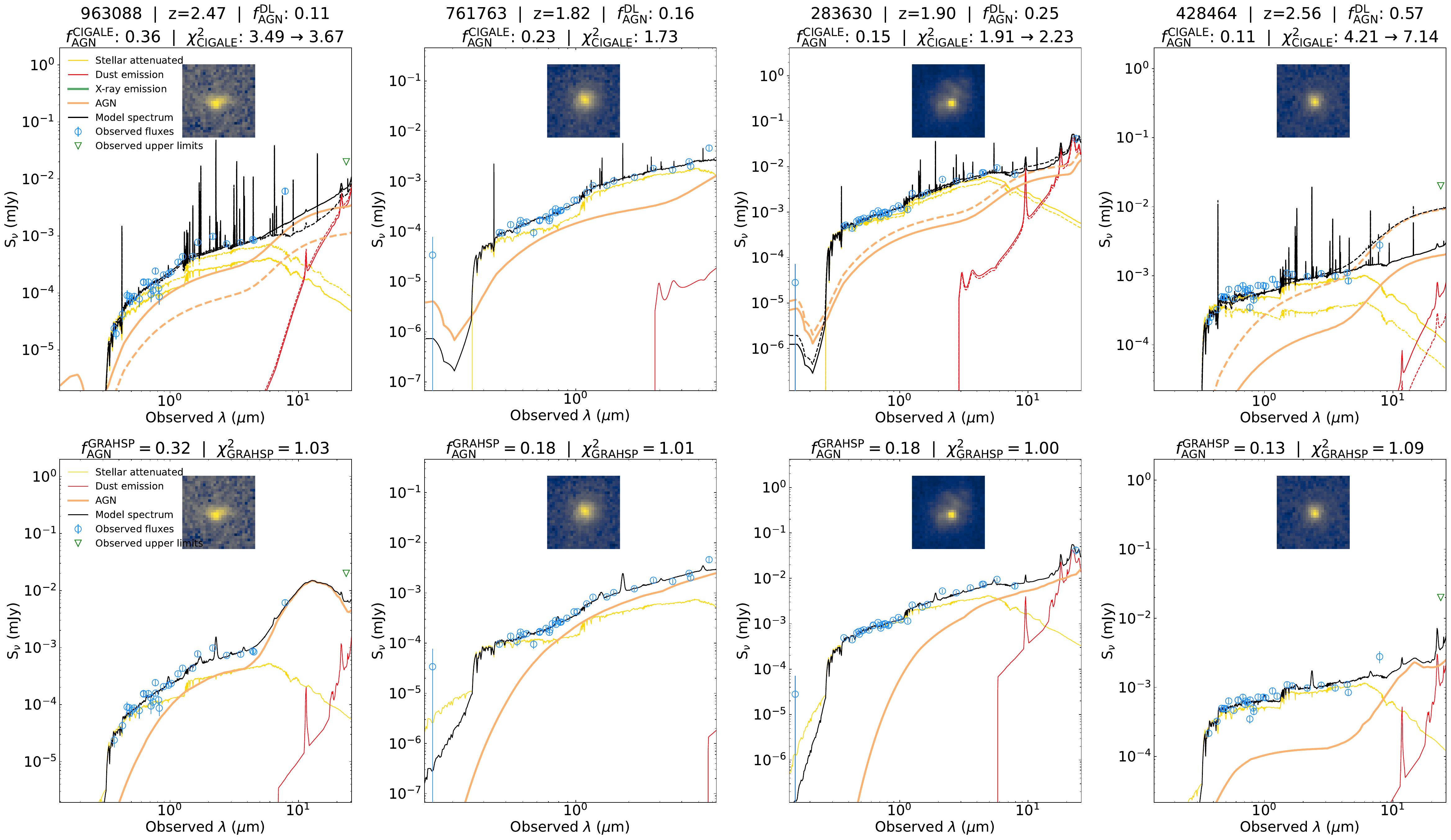}
\caption{
SED-fitting results and corresponding JWST/NIRCam F150W imaging for the four AGN-significant galaxies.
The $28\times28$ pixel cutouts ($\sim0.87^{\prime\prime}\times0.87^{\prime\prime}$ field of view) show the observed morphology for each source.
The first row presents the CIGALE fits, while the second row shows the GRAHSP results. For the three sources exhibiting significant scatter between DL prediction and SED decompositions, the dashed lines represent the refitted CIGALE results using a fixed $f_{\mathrm{AGN}}$ from DL. The resulting change in reduced $\chi^2$ is indicated in each subplot title.
}
\label{fig: AGN}
\end{figure*}

\begin{figure*}[h]
\centering
\includegraphics[width=0.975\textwidth]{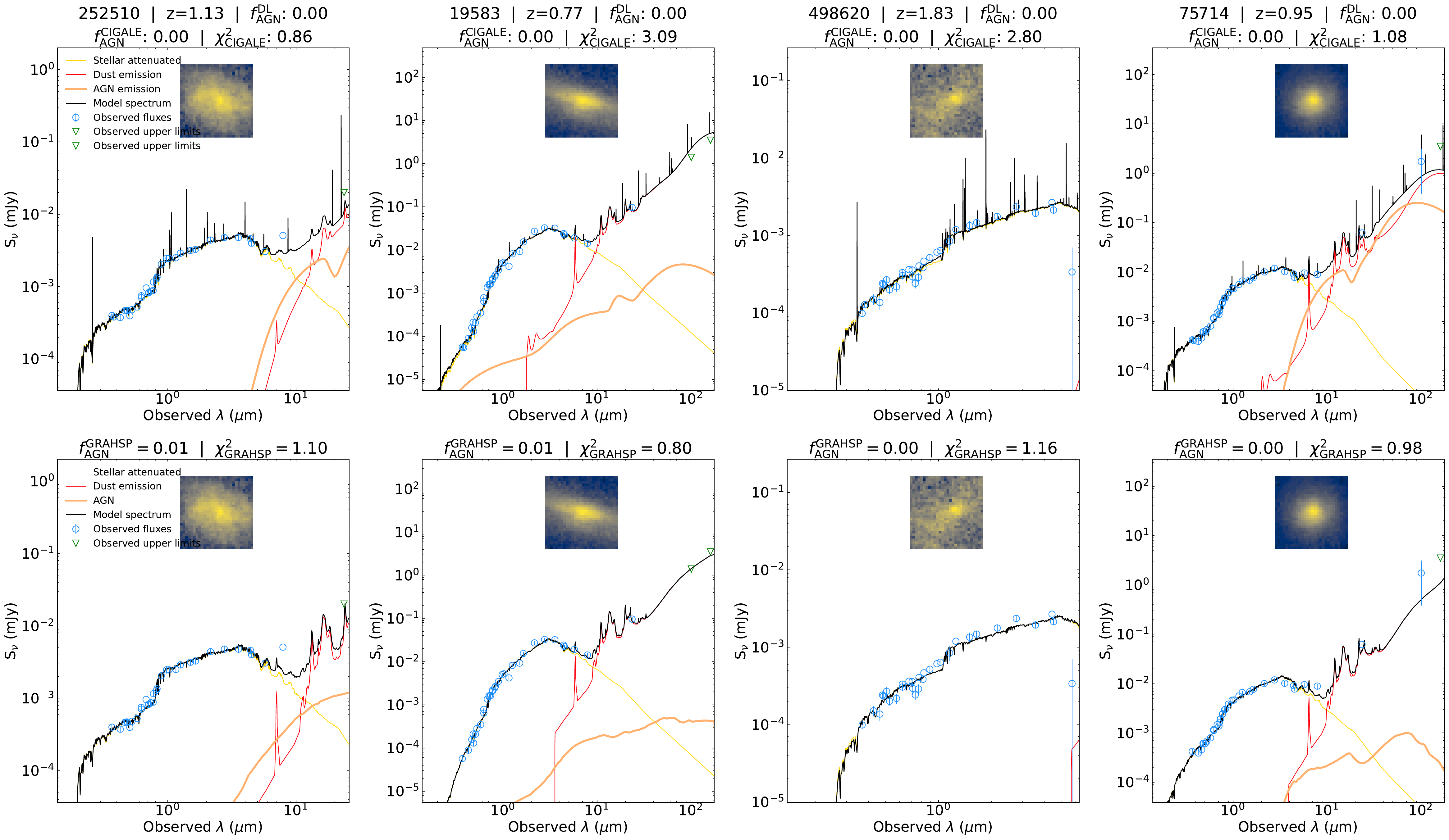}
\caption{SED-fitting results shown together with galaxy images for four randomly selected AGN-negligible galaxies, using the same filter, cutout size, and pixel scale as in Fig.~\ref{fig: AGN}.
}
\label{fig: noAGN}
\end{figure*}

Figure~\ref{fig: AGN} presents the four randomly selected example galaxies that are identified as AGN-significant by the DL prediction and both SED decompositions. Galaxies with IDs 963088, 283630, and 428464 exhibit relatively larger scatter in $f_{\mathrm{AGN}}^{\mathrm{DL}}$ compared to $f_{\mathrm{AGN}}^{\mathrm{SED}}$, and therefore undergo refitting with a fixed $f_{\mathrm{AGN}}^{\mathrm{DL}}$ value as described in Sect~\ref{refit}. The change in reduced $\chi^2$ is indicated in the titles of each sub-plot, and the dashed lines denote the refitted SED results. 
Figure~\ref{fig: noAGN} shows four randomly selected galaxies from the subsample with negligible AGN contributions according to the DL prediction and both SED decompositions.

\renewcommand{\arraystretch}{1.4}
\setlength{\tabcolsep}{3pt}
\setlength{\LTcapwidth}{\textwidth}
\begin{longtable}{c c c cc cc cc cc cc}
\caption{Galaxy properties of 30 AGN-significant galaxies identified by all three methods.}
\label{tab:galaxy_properties}\\

\hline
ID & Redshift & $f_{\mathrm{AGN}}^{\mathrm{DL}}$ & 
\multicolumn{2}{c}{$f_{\mathrm{AGN}}$} &
\multicolumn{2}{c}{$\log(\mathrm{SFR})$} &
\multicolumn{2}{c}{$\log(M_\star)$} &
\multicolumn{2}{c}{$A_V$} & 
\multicolumn{2}{c}{$\chi^2$} \\
 &  &  & 
CIGALE & GRAHSP &
C$^\dagger$ & G &
C & G &
C & G &
C & G \\
\hline
\endfirsthead

\hline
395859 & 1.4765 & 0.07 & 0.19 & 0.39 & 1.13 (1.06) & 0.50 & 10.35 (10.16) & 10.21 & 0.50 (0.50) & 0.17 & 1.16 (1.46) & 0.86 \\
30389 & 1.9102 & 0.06 & 0.25 & 0.13 & 0.83 (1.00) & 0.77 & 9.77 (9.74) & 9.82 & 0.10 (0.50) & 0.17 & 1.73 (2.50) & 0.97 \\
240330 & 1.9109 & 0.08 & 0.13 & 0.63 & 1.14 & -0.28 & 11.52 & 9.49 & 2.00 & 0.12 & 2.74 & 0.94 \\
283630 & 1.8953 & 0.25 & 0.15 & 0.18 & 1.78 (1.52) & 1.18 & 10.64 (10.28) & 10.15 & 0.50 (0.50) & 0.33 & 1.91 (2.23) & 1.00 \\
761763 & 1.8189 & 0.16 & 0.23 & 0.18 & 0.43 & 0.37 & 10.10 & 9.85 & 0.50 & 0.30 & 1.73 & 1.01 \\
796000 & 1.7552 & 0.82 & 0.24 & 0.24 & 0.39 (-0.14) & 0.28 & 9.32 (8.05) & 9.31 & 0.10 (0.05) & 0.13 & 1.89 (6.27) & 0.87 \\
122301 & 2.4652 & 0.65 & 0.82 & 0.62 & 1.73 (2.39) & 0.55 & 11.40 (10.95) & 8.52 & 5.00 (5.00) & 0.14 & 13.24 (13.88) & 1.25 \\
528974 & 2.1980 & 0.11 & 0.21 & 0.81 & 2.51 (2.19) & 0.10 & 9.42 (9.32) & 7.84 & 0.50 (0.50) & 0.24 & 3.03 (3.12) & 1.11 \\
543426 & 2.2097 & 0.16 & 0.36 & 0.45 & 0.31 (0.59) & 0.40 & 9.98 (9.73) & 9.19 & 0.50 (0.50) & 0.33 & 2.98 (3.20) & 1.16 \\
576291 & 2.0406 & 0.30 & 0.30 & 0.47 & -0.25 & -0.62 & 8.41 & 7.86 & 0.10 & 0.11 & 2.87 & 0.91 \\
681788 & 2.0344 & 0.54 & 0.35 & 0.37 & 1.43 (1.75) & -0.18 & 11.03 (10.96) & 9.70 & 3.00 (4.00) & 0.16 & 6.99 (7.80) & 1.05 \\
718056 & 2.1716 & 0.50 & 0.51 & 0.35 & 0.04 & -0.06 & 9.07 & 8.65 & 0.50 & 0.34 & 6.53 & 1.15 \\
798860 & 2.1224 & 0.10 & 0.29 & 0.77 & 0.45 (1.26) & -0.11 & 10.83 (10.47) & 8.26 & 2.00 (2.00) & 0.29 & 3.86 (4.23) & 0.98 \\
929188 & 2.1948 & 0.23 & 0.14 & 0.81 & 2.33 & -0.78 & 9.53 & 7.72 & 0.50 & 0.19 & 3.42 & 1.31 \\
945832 & 2.3367 & 0.07 & 0.22 & 0.75 & 0.73 (0.63) & -0.62 & 9.96 (9.77) & 8.34 & 0.50 (0.50) & 0.16 & 5.27 (5.42) & 1.08 \\
963088 & 2.4722 & 0.11 & 0.36 & 0.32 & 2.57 (2.39) & 0.52 & 9.48 (9.51) & 9.27 & 0.50 (0.50) & 0.39 & 3.49 (3.67) & 1.03 \\
21717 & 2.5885 & 0.30 & 0.30 & 0.86 & 1.78 & 0.22 & 11.45 & 8.63 & 3.00 & 0.23 & 2.96 & 0.87 \\
33237 & 2.6055 & 0.08 & 0.27 & 0.26 & 1.17 (1.16) & 0.68 & 10.22 (10.04) & 10.08 & 0.50 (0.50) & 0.21 & 1.60 (1.81) & 1.13 \\
65362 & 2.7969 & 0.14 & 0.15 & 0.90 & 1.34 & -1.06 & 11.01 & 7.31 & 2.00 & 0.13 & 3.57 & 1.19 \\
126847 & 2.9999 & 0.32 & 0.24 & 0.94 & 3.82 & -0.60 & 11.01 & 8.06 & 3.00 & 0.14 & 2.01 & 1.53 \\
139659 & 2.6735 & 0.34 & 0.41 & 0.48 & 1.19 & -0.23 & 10.86 & 9.06 & 3.00 & 0.33 & 2.87 & 1.82 \\
142501 & 2.5026 & 0.33 & 0.26 & 0.19 & 0.97 & 0.73 & 9.43 & 9.50 & 0.10 & 0.13 & 2.33 & 1.22 \\
317349 & 2.6649 & 0.55 & 0.18 & 0.26 & 1.15 (0.68) & 1.09 & 10.82 (10.25) & 10.29 & 0.10 (0.05) & 0.09 & 11.91 (13.43) & 1.34 \\
428464 & 2.5650 & 0.57 & 0.11 & 0.13 & 1.19 (0.79) & 0.98 & 9.65 (8.99) & 9.65 & 0.10 (0.05) & 0.13 & 4.21 (7.14) & 1.09 \\
435683 & 2.8470 & 0.32 & 0.46 & 0.76 & 2.65 (2.99) & 0.14 & 11.97 (11.67) & 8.73 & 5.00 (5.00) & 0.55 & 4.35 (4.04) & 1.36 \\
488530 & 2.6137 & 0.19 & 0.31 & 0.43 & 0.90 (0.73) & 0.46 & 9.37 (9.31) & 9.18 & 0.10 (0.10) & 0.07 & 3.72 (3.89) & 1.27 \\
625606 & 2.7213 & 0.10 & 0.50 & 0.53 & 1.98 (2.12) & 0.55 & 11.29 (10.85) & 9.66 & 3.00 (2.00) & 0.39 & 2.89 (3.00) & 0.88 \\
778992 & 2.7837 & 0.32 & 0.64 & 0.59 & 3.87 (3.50) & -0.08 & 10.47 (10.04) & 8.25 & 3.00 (1.50) & 0.08 & 10.87 (11.56) & 1.16 \\
852958 & 2.7143 & 0.47 & 0.26 & 0.22 & 0.20 (-0.16) & 0.57 & 9.02 (8.57) & 8.40 & 0.10 (0.05) & 0.07 & 8.73 (8.58) & 1.02 \\
934781 & 2.9699 & 0.45 & 0.81 & 0.61 & 1.96 (2.33) & 0.47 & 11.63 (11.06) & 9.10 & 4.00 (3.00) & 0.26 & 6.88 (8.09) & 1.16 \\

\hline

\multicolumn{13}{l}{\footnotesize $\dagger$ Values in parentheses indicate the refitted value.} \\

\end{longtable}

\end{appendix}

\end{document}